\documentclass[journal=jctcce,manuscript=article]{achemso}

\usepackage{algorithm2e}
\usepackage{amssymb,amsmath,multicol,multirow,longtable,array,mathpazo}
\usepackage{mhchem}
\usepackage{makecell}
\usepackage{enumerate}
\usepackage[T1]{fontenc} 

\usepackage{threeparttable}
\usepackage{color}
\usepackage{graphicx}
\usepackage{longtable}
\usepackage{enumerate}
\usepackage{mathrsfs}
\usepackage{amsfonts}
\usepackage{amsmath}
\usepackage{amssymb}
\usepackage{stmaryrd}
\usepackage{dcolumn}
\usepackage{bm}
\usepackage{bbm}
\usepackage{subfigure}
\usepackage{pgfplots}
\usepackage{verbatim}
\usepackage{accents}
\usepackage{diagbox}
\usepackage{simplewick} 

\usepackage[normalem]{ulem}
\usepackage{xcolor}
\newcommand\redsout{\bgroup\markoverwith{\textcolor{red}{\rule[0.5ex]{2pt}{0.4pt}}}\ULon}

\title{SDSPT2s: SDSPT2 with Selection}

\author{Yibo Lei}
\affiliation{Key Laboratory of Synthetic and Natural Functional Molecule of the Ministry of Education, College of Chemistry and Materials Science, Shaanxi key Laboratory of Physico-Inorganic Chemistry, Northwest University, Xi'an 710127, China}
\email{leiyb@nwu.edu.cn}

\author{Yang Guo}
\affiliation{Qingdao Institute for Theoretical and Computational Sciences, School of Chemistry and Chemical Engineering, Shandong University, Qingdao 266237, China}

\author{Bingbing Suo}
\affiliation{Institute of Modern Physics, Northwest University, and Shaanxi Key Laboratory for Theoretical Physics Frontiers, Xi'an 710127, China}

\author{Wenjian Liu}
\affiliation{Qingdao Institute for Theoretical and Computational Sciences, School of Chemistry and Chemical Engineering, Shandong University, Qingdao 266237, China}
\email{liuwj@sdu.edu.cn}

\begin{document}

\begin{abstract}
As an approximation to SDSCI [static-dynamic-static (SDS) configuration interaction (CI), a minimal MRCI; Theor. Chem. Acc. 133, 1481 (2014)], SDSPT2 [Mol. Phys. 115, 2696 (2017)]
is a CI-like multireference (MR) second-order perturbation theory (PT2)
that treats single and multiple roots in the same manner. This feature permits the use of configuration selection
over a large complete active space (CAS) $P$ to end up with a much reduced reference space $\tilde{P}$,
which is connected only with a small portion ($\tilde{Q}_1$) of the full first-order interacting space $Q$ connected to $P$.
The most expensive portion of the reduced interacting $\tilde{Q}_1$ space (which involves three active orbitals)
can further be truncated by partially bypassing its generation followed by
an integral-based cutoff.
With marginal loss of accuracy, the selection-truncation procedure, along with an efficient evaluation and storage of internal contraction coefficients,
renders SDSPT2s (SDSPT2 with selection) applicable to systems that cannot be handled by the parent CAS-based SDSPT2,
as demonstrated by several challenging showcases.
\end{abstract}

\maketitle

\clearpage
\newpage

\section{Introduction}\label{sec:intro}
A system of strongly correlated electrons is characterized by a large number of energetically adjacent and singly occupied frontier orbitals.
The strong static/nondynamic correlation among such orbitals renders the many-electron wave function
a heavy mixture of a huge number of Slater determinants or
equivalently configuration state functions (CSF). As such, such systems (especially of low spins) go beyond the capability of
single reference methods. Instead, the use of a multireference (MR) method is mandatory. The available MR methods can be classified into
three families\cite{iCI}: static-then-dynamic (SD), dynamic-then-static (DS), and static-dynamic-static (SDS).
The SD family of methods start with a diagonalization of the bare molecular Hamiltonian projected onto
a reference/active space $P=\{\Phi_{R}; R=[1,N_R]\}$, so as to
obtain a reference state $\Psi^{(0)}_k=\sum_{R=1}^{N_R}\Phi_{R} \bar{C}_{R k}^{(0)}$.
This step captures static correlation, especially when $\Psi^{(0)}_k$ is only qualitatively or semi-quantitatively correct.
The remaining dynamic correction to $\Psi^{(0)}_k$ can be accounted for in a number of ways, even just to first order $\Psi^{(1)}_k$.
Classic examples of this family of methods include complete active space (CAS) second-order perturbation theory (CASPT2)\cite{CASPT2a,CASPT2b},
multiconfigurtion quasi-degenerate perturbation theory\cite{MCQDPT2a}, and n-electron valence
second-order perturbation theory (NEVPT2)\cite{NEVPT2a,NEVPT2b}.
A common feature of such methods lies in that the coefficients $\bar{C}_{R k}^{(0)}$ of the reference state $\Psi^{(0)}_k$ are used to construct $\Psi^{(1)}_k$
for dynamic correlation but are not relaxed in the presence of dynamic correlation. Colloquially,
dynamic correlation sees static correlation but static correlation
does not see dynamic correlation. This is true even for their multi-state (MS) variants\cite{MSCASPT2,XMCQDPT2,XMSCASPT2,MSNEVPT2}
because of their insufficient revision of the coefficients $\bar{C}_{R k}^{(0)}$.
In contrast, the DS family of methods incorporate dynamic correction to each reference function $\Phi_{R}$ and then
construct and diagonalize an effective Hamiltonian in the space $P$, thereby producing a wave function that can be (very) different from $\Psi^{(0)}_k$
obtained by diagonalizing the bare Hamiltonian in the same space. Classic examples
are the (shifted) B$_{\mathrm{k}}$ type of methods\cite{BWBka,BWBkb,BWBkc,RSBka,DPD,DCD-CAS2}.
Since the zero-order coefficients $\bar{C}_{R k}^{(0)}$ are not involved in the dynamic correlation step at all,
it can be said that dynamic correlation does not see static correlation in such family of methods.
It should hence be clear that both the SD and DS families of methods cannot achieve balanced treatments of the static and dynamic components of
the overall correlation, especially when the two components are strongly entangled and even interchangeable.
This situation warrants a SDS type of treatment, where the predetermined coefficients $\bar{C}_{R k}^{(0)}$ are used
in dynamic correlation but are then sufficiently or even fully relaxed in the presence of dynamic correlation.
That is, the static and dynamic components of the overall correlation do see each other.
Several variants of multireference second-order perturbation theory (MRPT2)\cite{GVVPT2a,GVVPT2b,GVVPT2c,SS-MRPT2a,SS-MRPT2b,SS-MRPT2c,SS-MRPT2d,SS-MRPT2e,DSRGPT2a,DSRGPT2b},
internally contracted multireference configuration interaction (ic-MRCI)
\cite{icMRCI1c,icMRCI2a0,icMRCI2a,icMRCI2b,icMRCI3a,Wang2003CICPL,WenicMRCI,XianCI2018,ORCAicMRCI,SongGUGA2024,RR-MRCI,MRCIRev,MCSCFrev2012},
and internally contracted multireference coupled-cluster methods
\cite{icMRCCa,SS-MRexpT,icMRCCb,icMRCCc,icMRCCd,icMRCCe,icMRCCf,icMRCCg,SS-MRCC1,SS-MRCC2,SS-MRCCa,SS-MRCCb,Malrieu-MS-MRCC,MRCCRev}
belong to this family. However, such methods usually require the construction and diagonalization of a large Hamiltonian matrix
even just for one state.
At variance with this, one of the present authors proposed\cite{SDS} a restricted SDS framework for
constructing many-electron wave functions, where no matter how many electrons and how many orbitals are to be correlated,
only a $3N_P$-by-$3N_P$ Hamiltonian matrix is constructed and diagonalized for $N_P$ states.
The framework leads to a series of methods, including SDSCI\cite{SDS}, SDSPT2\cite{SDSPT2}, iterative configuration interaction (iCI)\cite{iCI},
iCI with configuration and perturbation (iCIPT2)\cite{iCIPT2,iCIPT2New}, and extended variants\cite{LiuWIRES2023} of SDSCI and SDSPT2. Albeit a minimal MRCI, SDSCI
is very close in accuracy to ic-MRCI\cite{SDSrev,SDSradical}, with a computational cost being only that
of one iteration of ic-MRCI. As an approximation to SDSCI, SDSPT2 is a CI-like MRPT2
and treats single and multiple roots in the same way.
In particular, SDSPT2 gives rise to MS-NEVPT2 for free. While SDSPT2 is usually very similar to MS-NEVPT2 in accuracy
\cite{SDSrev,SDSradical,SDStest1,SDStest2,SDStest3,SDStest4}, it does outperform MS-NEVPT2
for situations with multiple nearly degenerate states\cite{SDSPT2} .
As an iterative version of SDSCI, iCI is an exact solver of full CI, whereas iCIPT2 is one of the most efficient
near-exact methods\cite{benzene}. Taking iCI as the CASCI solver, we obtain iCISCF\cite{iCISCF}, which can handle
active spaces as large as (60e, 60o).

One major problem associated with the above multireference methods lies in that a large reference space $P$
leads to an exceedingly large first-order interacting space (FOIS) $Q$  that is intractable.
The only way to go is to reduce the reference space from $P$ to $\tilde{P}$,
so as to reduce the FOIS from $Q$ to $\tilde{Q}_1$.
The $\tilde{Q}_1$ space can further be treated approximately by separating it into an important subset
that is treated rigorously and an unimportant subset that can be treated approximately.
Several approaches have been proposed along this line
\cite{RMS-MRPT2,selectMRPT2,GMSQDPT2,selectMRMP2}, which differ from
each other in the choice of reduced reference space $\tilde{P}$, zeroth-order Hamiltonian $H_0$, perturbers spanning $\tilde{Q}_1$,
and effective Hamiltonian to be diagonalized for the final solutions.
In the present work, we apply such approximations to the
complete active space self-consistent field (CASSCF)-based SDSPT2\cite{SDSPT2}, so as to render it
applicable to systems that require very large active spaces.

The paper is organized as follows. The essential features of SDSPT2\cite{SDSPT2} are first recapitulated in Sec. \ref{SDSPT2}. The
hole-particle symmetry-based graphic unitary group approach (HPS-GUGA)\cite{Wang2003CICPL} is then employed in Sec. \ref{implement} to
generate excited configuration state functions (CSF) from individual orbital configurations (oCFG) contained
in the reduced reference wave functions
$\tilde{\Psi}^{(0)}_k=\sum_{R=1}^{\tilde{N}_R}\Phi_{R} \tilde{\bar{C}}_{R k}^{(0)}$ that result from truncation
of the CASSCF/iCISCF\cite{iCISCF} wave functions $\Psi^{(0)}_k$ ($=\sum_{R}^{N_R}\Phi_{R}\bar{C}_{R k}^{(0)}$).
Noticeably, many of the so-generated CSFs do not belong to the $\tilde{Q}_1$ space and hence
have zero matrix elements with the reduced reference wave functions
$\tilde{\Psi}^{(0)}_k$. As such, they should first be screened out. Moreover, those double excitations involving three active orbitals (TAOD)
are computationally most expensive and will hence be screened during the construction of reduced sub-DRTs (distinct row table\cite{Shavitt1977})
and also by an integral-based cutoff.
The efficacy of the resulting SDSPT2s (SDSPT2 with selection) is illustrated in Sec. \ref{Results} with
several challenging systems, including a simplified model of heme, chromium dimer, $\textrm{Cu$_2$O$^{2+}_2$}$ core, and cobalt nitrosyl complex [Co(TC-3,3)(NO)]. The paper is closed with a summary in Sec. \ref{Conclusion}.

Unless otherwise stated, the notations documented in Table \ref{Notation} are to be used for the orbitals and states, under the Einstein summation
convention over repeated indices.

\begin{threeparttable}
  \caption{Notations for the orbitals and states}
  \tabcolsep=20pt
  \scriptsize
  \begin{tabular}{p{170pt}p{70pt}p{100pt}}
   \hline\hline
   space & & indices\\
   \hline
         &orbital space & \\
   arbitrary orbitals & &$p,q,r,s$ \\
   hole ( or closed) orbitals & &$i,j$ \\
   active orbitals & &$u,v,t,w$ \\
   external ( or virtual) orbitals  & & $a,b$ \\
         &   configuration space & \\
   step vector in HPS-GUGA && $(\mathbf{d})_\mu=[(\mathbf{d})_e(\mathbf{d})_a(\mathbf{d})_h]_\mu$ \\
   configuration state function (CSF)  && $\Phi_{\mu}$ \\
   reference CSF && $\Phi_R$ \\
   reference orbital configuration && $\Phi_{\bar{R}}^{\mathrm{o}}=\Pi_{i=1}^n\psi_i^{n_i}$; $n_i\in[0,2]$ \\
   excited CSF && $\Phi_q$ \\
   reference wavefunction && $\Psi^{(0)}_k $ \\
   electronic state && $\Psi_I$ \\
   node of distinct row tableaux (DRT) && $(a_r,b_r)_J$ \\
   sub-DRT && $(\bar{\mathrm{X}}\mathrm{Y})$ \\
   internally contracted configuration && $\bar{\Phi}^{\bar{\mathrm{X}}\mathrm{Y}}_{Mk}$ \\
   complete active space && $P$\\
   full first-order interacting space (FOIS)&& $Q$\\
   reduced reference space&& $\tilde{p}$\\
   reduced FOIS generated from $\{\Phi_{\bar{R}}^{\mathrm{o}}\}$ in $\tilde{P}$&& $\tilde{Q}=\tilde{Q}_1+\tilde{Q}_r$\\
   reduced FOIS interacting with $\tilde{P}$&& $\tilde{Q}_1$\\
   \hline \hline
  \end{tabular}\label{Notation}
\end{threeparttable}

\section{SDSPT2}\label{SDSPT2}
The \emph{restricted} SDS framework\cite{SDS} starts with the following wave functions for $N_P$ lowest states
\begin{eqnarray}
|\Psi_I\rangle
=\sum_k^{N_P} |\Psi_k^{(0)}\rangle \tilde{C}_{kI} + \sum_k^{N_P}|\Xi_k^{(1)}\rangle \tilde{C}_{(k+N_P)I} + \sum_k^{N_P}|\Theta_k^{(2)}\rangle \tilde{C}_{(k+2N_P)I},\label{WF}
\end{eqnarray}
where $\Psi_k^{(0)}$ ($=\sum^{N_R}_{R=1}\Phi_R \bar{C}_{{R}k}^{(0)}$), $\Xi_k^{(1)}$, and $\Theta_k^{(2)}$ represent the zeroth-order, first-order, and secondary functions, respectively. By introducing the primary ($P_m$) and secondary ($P_s$) parts of the $P$ space,
\begin{eqnarray}
P_m&=&\sum_{k=1}^{N_P}|\Psi^{(0)}_k\rangle\langle\Psi_k^{(0)}|,\label{Pmdef}\\
P_s&=&P-P_m=\sum_{l=N_P+1}^{N_R}|\Psi^{(0)}_l\rangle\langle\Psi_l^{(0)}|,\label{Psdef}
\end{eqnarray}
the first-order ($\Xi^{(1)}_k$) and secondary ($\Theta^{(2)}_k$) functions can be defined as
\begin{eqnarray}
|\Xi^{(1)}_k\rangle &=& Q\frac{1}{E_k^{(0)}-H_0}QH|\Psi^{(0)}_k\rangle=\sum_{q\in Q}|\Psi^{(0)}_{q,k}\rangle\bar{C}^{(1)}_{qk},\label{PTWF1def}\\
Q &=& 1-P=\sum_{q\in Q}|\Psi^{(0)}_{q,k}\rangle\langle\Psi^{(0)}_{q,k}|,\label{Q1def}\\
|\Theta^{(2)}_k\rangle &=& P_sH|\Xi^{(1)}_k\rangle \label{LanczosVec}\\
&\approx& P_s^\prime H|\Xi^{(1)}_k\rangle,\quad P_s^\prime=\sum_{l=N_P+1}^{N_P+M_P}|\Psi^{(0)}_l\rangle\langle\Psi_l^{(0)}|,\label{Psadef}\\
&=&\sum_{R\in P}|\Phi_R\rangle\bar{C}^{(2)}_{Rk}\label{SecondaryWFdef},\\
\bar{C}^{(2)}_{Rk}&=&\sum_{S\in P}
\sum_{q\in Q}\bar{D}^{(0)}_{RS}\langle\Phi_S|H|\Psi_{q,k}^{(0)}\rangle\bar{C}^{(1)}_{qk},\label{C2coff}\\
\bar{D}^{(0)}_{RS}&=&\sum_{l=N_P+1}^{N_P+M_P}\bar{C}_{Rl}^{(0)} \bar{C}_{Sl}^{(0)},\quad R, S\in P.
\end{eqnarray}
Note that $\{|\Psi^{(0)}_{q,k}\rangle\}$ in Eq. \eqref{PTWF1def} are eigenvectors of a zeroth-order Hamiltonian $H_0$,
\begin{eqnarray}
H_0|\Psi^{(0)}_{q,k}\rangle=E_{q,k}^{(0)}|\Psi^{(0)}_{q,k}\rangle,\quad q\in Q,\quad k\in[1,N_P],\label{DyallDiag}
\end{eqnarray}
which may be state-specific or state-collective (independent of state $k$).
Eq. \eqref{LanczosVec} represents the projection of the Lanczos vector $H|\Xi^{(1)}_k\rangle$ onto the secondary space,
which is further approximated to Eq. \eqref{Psadef} to simplify the evaluation of the matrix elements.
It is clear from Eq. \eqref{SecondaryWFdef} that the secondary functions are linear combinations
of the reference CSFs $\{\Phi_R\}$ but with their coefficients related to the first-order functions.
As such, they can facilitate the relaxation of the reference coefficients $\bar{C}^{(0)}_{R k}$ in the presence of dynamic correlation,
particularly for avoided crossings\cite{iCI} or multiple near-degenerate states\cite{SDSPT2}.
The fact that $|\Psi_k^{(0)}\rangle$, $|\Xi_k^{(1)}\rangle$, and $|\Theta_k^{(2)}\rangle$ have decreasing weights
in the wave function $|\Psi_I\rangle$ justifies the characterization of them as primary, external, and secondary states.
Since both $|\Xi_k^{(1)}\rangle$ and $|\Theta_k^{(2)}\rangle$ are specific to $|\Psi_k^{(0)}\rangle$, the generalized eigenvalue problem
\begin{eqnarray}
\tilde{\mathbf{H}}\tilde{\mathbf{C}}=\tilde{\mathbf{S}}\tilde{\mathbf{C}}\tilde{\mathbf{E}}
\end{eqnarray}
for determining the expansion coefficients of $|\Psi_I\rangle$ is only of dimension $3N_P$. The Hamiltonian and metric matrices have the following structures
\begin{eqnarray}
\tilde{\textbf{H}} &=& \begin{pmatrix}
P_mHP_m & P_mHQ & P_mHP_s\\
QHP_m   & QHQ   & QHP_s  \\
P_sHP_m & P_sHQ &P_sHP_s\end{pmatrix}\label{Hmat1}\\
&=& \begin{pmatrix}
 E_{k}^{(0)}\delta_{kl} & \langle\Psi_k^{(0)}|H|\Xi_l^{(1)}\rangle &0 \\
 \langle\Xi_l^{(1)}|H|\Psi_k^{(0)}\rangle& \langle\Xi_k^{(1)}|H|\Xi_l^{(1)}\rangle &\langle\Xi_k^{(1)}|H|\Theta_l^{(2)}\rangle\\
 0&\langle\Theta_l^{(2)}|H|\Xi_k^{(1)}\rangle& \langle\Theta_k^{(2)}|H|\Theta_l^{(2)}\rangle\end{pmatrix},\ k,l=1,\cdots, N_P,\label{Hmat2}\\
\tilde{\textbf{S}}&=&\begin{pmatrix}\delta_{kl} & 0 &0 \\
 0& \langle\Xi_k^{(1)}|\Xi_l^{(1)}\rangle & 0\\
 0&0& \langle\Theta_k^{(2)}|\Theta_l^{(2)}\rangle\end{pmatrix}. \label{Smat}
\end{eqnarray}

The above is nothing but a minimal MRCI (dubbed as SDSCI\cite{SDS}). Taking SDSCI as the seed,
various methods can be derived\cite{iCI,iCIPT2,iCIPT2New,LiuWIRES2023}, among which the simplest variant
is SDSPT2, which amounts to replacing the $QHQ$ block in Eq. \eqref{Hmat1} with $QH_0Q$. Like NEVPT2\cite{NEVPT2a},
the Dyall CAS/A Hamiltonian\cite{DyallH0} is adopted here for $H_0$:
\begin{align}
&H_0=H_I^D + H_A^D, \label{HDoper}\\
&H_I^D=\sum_{i}\epsilon_i\hat{E}_{ii} + \sum_{a}\varepsilon_a\hat{E}_{aa} +C_I^D,\quad \hat{E}_{pq}=a^{p\sigma}a_{q\sigma},\label{HIoper}\\
&H_A^D=\sum_{tu} f^c_{tu}\hat{E}_{tu}+\frac{1}{2}\sum_{tuvw}(tu|vw)\hat{e}_{tu,vw},\quad \hat{e}_{tu,vw}=\hat{E}_{tu}\hat{E}_{vw}-\delta_{uv}\hat{E}_{tw}, \label{HAoper}\\
&C_I^D= E^c-2\sum_{i}\varepsilon_i,\ E^c=\sum_{i}(h_{ii}+f^c_{ii}), \label{DyallC}\\
&f^c_{pq}=h_{pq}+\sum_{i}[2(pq|ii)-(pi|iq)],
\end{align}
where the quasi-canonical orbital energies, $\epsilon_i$ and $\epsilon_a$,
are obtained by diagonalizing the generalized Fock matrix
\begin{align}
&F_{pq}=f^c_{pq}+\sum_{tu}[(pq|tu)-\frac{1}{2} (pu|tq)]D_{tu}, \\
&D_{tu}=\sum_{k} w_k\langle\Psi_k^{(0)}|\hat{E}_{tu}|\Psi_k^{(0)}\rangle
\end{align}
for the doubly occupied and virtual subspaces separately.
The particular choice of the constant $C^D_{\mathrm{I}}$ \eqref{DyallC} is to
make $\langle\Phi_{\mu}|H^D|\Phi_{\nu}\rangle$ equal to $\langle\Phi_{\mu}|H|\Phi_{\nu}\rangle$  for all $\Phi_R$
in $P$, such that
\begin{eqnarray}
\langle \Psi_k^{(0)}|H^D|\Psi_l^{(0)}\rangle=\langle \Psi_k^{(0)}|H|\Psi_l^{(0)}\rangle=E_k^{(0)}\delta_{kl},\label{E1D}
\end{eqnarray}
where
\begin{eqnarray}
E_k^{(0)}=E^c+\sum_{tu}f^c_{tu}\langle \Psi^{(0)}_k|\hat{E}_{tu}|\Psi^{(0)}_k\rangle+\frac{1}{2}\sum_{tuvw}(tu|vw)\langle \Psi^{(0)}_k|\hat{e}_{tu,vw}|\Psi^{(0)}_k\rangle.
\end{eqnarray}

Since $H_0$ \eqref{HDoper} does not couple\cite{SDSPT2} different classes of excitations (cf. Table \ref{sub-DRT}),
Eq. \eqref{DyallDiag} can be solved for each class ($K\in[1,8]$) of excitations separately. Still, however,
this is possible only when the state-specific, orthonormalized internally contracted configurations (ICC)\cite{selectMRPT2}
are used as the basis to expand the zeroth-order eigenvectors (perturbers) $|\Psi^{(0)}_{(qK),k}\rangle$ (vide post). Once this is done,
the $QH_0Q$ block in Eq. \eqref{Hmat1} can readily be constructed as
\begin{eqnarray}
\langle \Xi_{k}^{(1)}|H_0|\Xi_{l}^{(1)}\rangle&=&\sum_{(qK),(rK)\in Q}
\bar{C}_{(qK),k}^{(1)}\langle\Psi_{(qK),k}^{(0)}|H_0|\Psi_{(rK),l}^{(0)}\rangle\bar{C}_{(rK),l}^{(1)}\\
&\approx&\frac{1}{2}
\sum_{(qK),(rK)\in Q} [E_{(qK),k}^{(0)}+E_{(rK),l}^{(0)}]
 \langle\Psi_{(qK),k}^{(0)}|\Psi_{(rK),l}^{(0)}\rangle
 \bar{C}_{(qK),k}^{(1)}\bar{C}_{(rK),l}^{(1)},\label{HDmat}\\
\bar{C}_{(qK),k}^{(1)}&=&\frac{\langle\Psi^{(0)}_{(qK),k}|H|\Psi_k^{(0)}\rangle}{E_k^{(0)}-E_{(qK),k}^{(0)}},\quad K\in[1,8].\label{C1coff}
\end{eqnarray}
Since the most numerous matrix elements $\langle\Psi_{(qK),k}^{(0)}|H|\Phi_R\rangle$ are shared by $\bar{C}_{(qK),k}^{(1)}$ \eqref{C1coff},
$\bar{C}_{(qK),k}^{(2)}$ \eqref{C2coff}, $QHP_m$/$P_mHQ$, and $QHP_s$/$P_sHQ$, and the other matrix elements in Eqs. \eqref{Hmat2}
and \eqref{Smat} can readily be evaluated, SDSPT2 has little computational overhead over MS-NEVPT2
(which only requires the $P_mHP_m$ and $P_mHQ$ blocks). It is obvious that SDSPT2 can produce MS-NEVPT2 for free.
However, it should be noted that, unlike NEVPT2, SDSPT2 is not size consistent. Nevertheless, the size consistency errors
can readily be cured by the Pople correction\cite{Size4}, as demonstrated before\cite{SDSrev}.

\begin{threeparttable}
  \caption{Correspondence between excitation operators $\hat{E}_M$ and sub-DRTs ($\bar{\mathrm{X}}\mathrm{Y}$)}
  \tabcolsep=30pt
  \scriptsize
  \begin{tabular}[h]{@{}ccccclccccccc@{}}
   \hline\hline
   Class $K$&  $\hat{E}_M$  & $N_d$ \tnote{a} & $N_e$ \tnote{b} & $\bar{\mathrm{X}}\mathrm{Y}$ \tnote{c} & $N^{ex}_{\bar{\mathrm{X}}\mathrm{Y}}$ \tnote{d} \\
   \hline
0& $\hat{E}_{uv},\hat{e}_{uv,tw}$   & 0 & 0 & $\bar{\mathrm{V}}\mathrm{V}$ & 0 \\
1& $\hat{E}_{ui}, \hat{e}_{ui,vw}$  & 1 & 0 & $\bar{\mathrm{D}}\mathrm{V}$ & 0, 1, 2 \\
2& $\hat{E}_{au}, \hat{e}_{au,vw}$  & 0 & 1 & $\bar{\mathrm{V}}\mathrm{D}$ & 0, 1 \\
3& $\hat{E}_{ai}, \hat{e}_{ai,uv}, \hat{e}_{ui,av}$
                                    & 1 & 1 & $\bar{\mathrm{D}}\mathrm{D}$ & 0, 1 \\
4& $\hat{e}_{ui,vj}$                & 2 & 0 & $\bar{\mathrm{P}}\mathrm{V}$ & 0, 1, 2 \\
5& $\hat{e}_{au,bv}$                & 0 & 2 & $\bar{\mathrm{V}}\mathrm{P}$ & 0 \\
6& $\hat{e}_{ai,uj}$                & 2 & 1 & $\bar{\mathrm{P}}\mathrm{D}$ & 0, 1 \\
7& $\hat{e}_{ai,bu}$                & 1 & 2 & $\bar{\mathrm{D}}\mathrm{P}$ & 0 \\
8& $\hat{e}_{ai,bj}$                & 2 & 2 & $\bar{\mathrm{P}}\mathrm{P}$ & 0 \\
   \hline \hline
  \end{tabular}\label{sub-DRT}
  \begin{tablenotes}
  \item[a]Number of holes in the hole space.
  \item[b]Number of electrons in the virtual space.
  \item[c]P = S or T; S: two-electron singlet; T: two-electron triplet; D: one-electron doublet; V: void.
  \item[d]Allowed excitation number of sub-DRT $(\bar{\mathrm{X}}\mathrm{Y})$.
  \end{tablenotes}
\end{threeparttable}
\section{Implementation of SDSPT2s}\label{implement}
\begin{figure}
\centering
\includegraphics[width=0.7\textwidth]{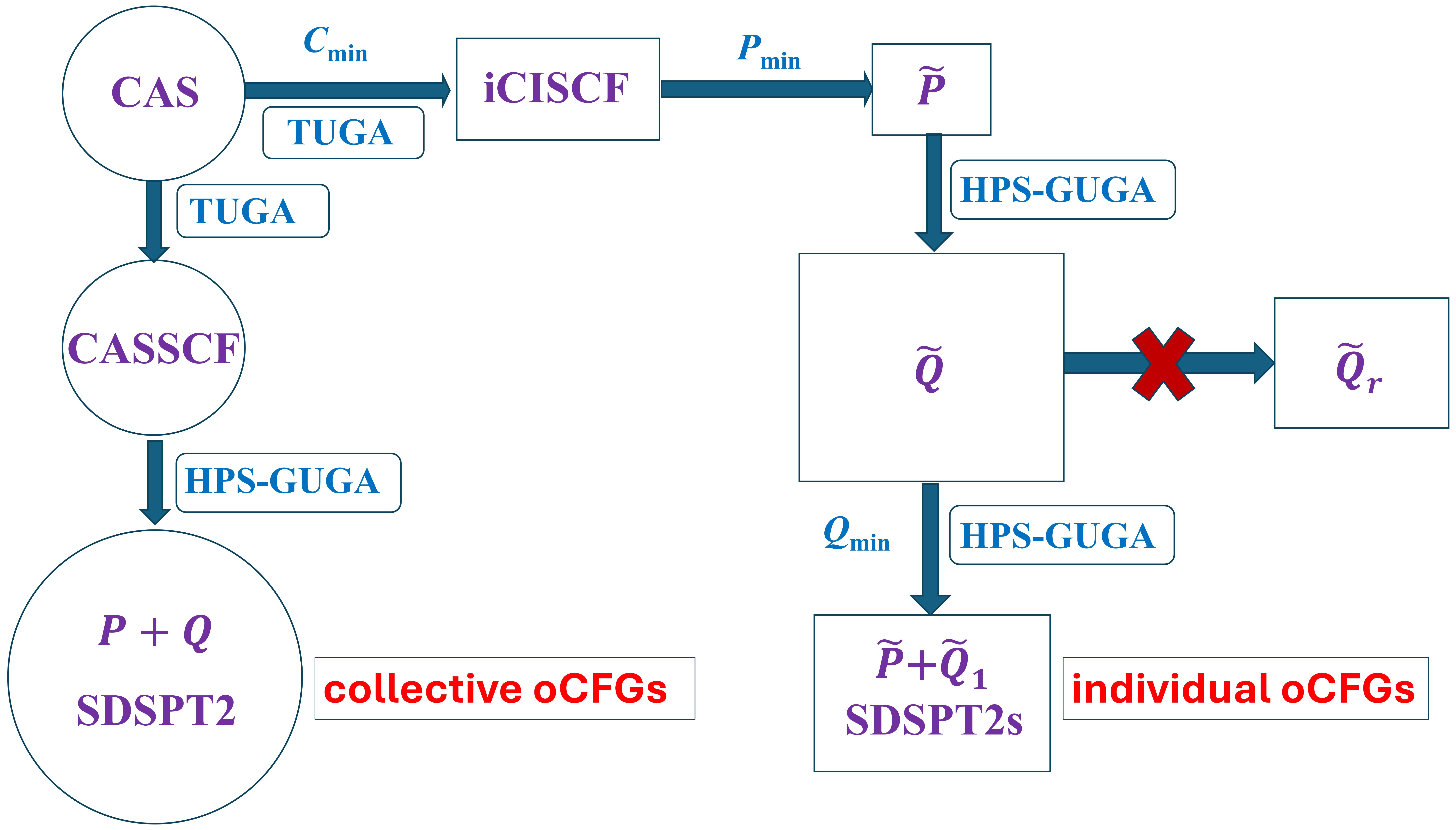}
\caption{CAS- and selection-based SDSPT2.
TUGA: tabulated unitary group approach\cite{iCIPT2} for basic coupling
coefficients between randomly selected CSFs. $\tilde{Q}_r$ ($=\tilde{Q}-\tilde{Q}_1$) does not interact with $\tilde{P}$.
Collective/individual oCFGs: collective/individual treatment of reference orbital configurations.}
\label{SDSPT2-scheme}
\end{figure}

Even though the FOIS $Q$ can be decomposed into 8 disjoint subspaces (see Table \ref{sub-DRT}),
the major computational step of the CAS-based SDSPT2\cite{SDSPT2} (and NEVPT2\cite{NEVPT2a,NEVPT2b,MSNEVPT2})
is still the diagonalization of the Dyall Hamiltonian \eqref{HDoper}
to produce the perturbers $\{\Psi_{(qK),k}^{(0)}\}$ \eqref{DyallDiag}.
To remedy this, the SDSPT2s scheme is proposed here.
The very first step is to reduce the CAS $P$ to $\tilde{P}$. That is, after the iCISCF(2)\cite{iCISCF} calculations
(NB: iCISCF(2) emphasizes that the total energy is the sum of the iCISCF variational energy and second-order Epstein-Nesbet perturbation
(ENPT2)\cite{Epstein,Nesbet}  correction within the CAS), only those CSFs with coefficients
larger in absolute value than the preset threshold $P_{\mathrm{min}}$ for any state are to be retained.
The survival CSFs $\{\Phi_R\}_{R=1}^{\tilde{N}_R}$ span the $\tilde{P}$ space. Although it is in principle possible to
generate the interacting excited CSFs $\Phi_q\in \tilde{Q}_1$ directly from $\{\Phi_R\}_{R=1}^{\tilde{N}_R}$, it is in practice more efficient\cite{selectMRPT2,ICEa} to
obtain them by exciting the reference orbital configurations (oCFG) $\{\Phi_R^{\mathrm{o}}\}_{R=1}^{N_R^{\mathrm{o}}}$ contained in $\tilde{P}$, via the HPS-GUGA\cite{Wang2003CICPL}.
A problem lies in that $\{\Phi_R^{\mathrm{o}}\}_{R=1}^{N_R^{\mathrm{o}}}$ are
usually just a subset of those oCFGs contained in space $P$, such that the HPS-GUGA\cite{Wang2003CICPL} designed for a CAS reference
must be modified (see Sec. \ref{CI-subspace}). Another problem that does not occur to a CAS reference lies in that
many excited CSFs $\Phi_r\in\tilde{Q}_r=\tilde{Q}-\tilde{Q}_1$ generated this way do not interact with the $\tilde{P}$ space.
Nevertheless, they can readily be eliminated by invoking
the conditions $\langle\Phi_r|\hat{E}_{pq}|\Psi_k^{(0)}\rangle=0$ and $\langle\Phi_r|\hat{e}_{pq,rs}|\Psi_k^{(0)}\rangle=0$.
Note in passing that some active orbitals of $P$ may disappear in $\tilde{P}$. Although
such situation should in principle be avoided (by reducing $P_{\mathrm{min}}$),
it does not really matter in practice, for their contributions are already contained in the iCISCF(2) energy.
For clarity, SDSPT2s is compared with the parent, CAS-based SDSPT2 in Fig. \ref{SDSPT2-scheme}.

\subsection{Generation of excited CSFs}\label{CI-subspace}
The $n$ spatial orbitals to be correlated are first partitioned into $n_d$ doubly occupied (DOS), $n_a$ active (AOS), and $n_e$ external (EOS) orbitals.
The former two together can be called internal orbitals (IOS). In the present context, the orbitals are arranged
as $(1,\cdots, n_e; n_e+1,\cdots, n_e+n_a; n_e+n_a+1, \cdots, n)$.
The reference oCFGs $\{\Phi_R^{\mathrm{o}}\}$
differ only in the occupation numbers $N_{w,R}^{\mathrm{o}}$ of the $n_a$ active orbitals.
It is certainly possible to first generate the unique set of singly and doubly
excited oCFGs $\{\Phi_q^{\mathrm{o}}\}$, from which the excited CSFs can then be obtained by using, e.g., the genealogical coupling scheme\cite{HelgakerBook}.
However, more efficient is the graphical unitary group approach, where the so-called DRT\cite{Shavitt1977}
can be constructed very efficiently for a compressed recording of the excited CSFs.
This scheme can further be simplified by observing\cite{Wang2003CICPL} that the partial DRTs for the EOS and DOS
can easily be established in advance, for they have at most two electrons and two holes, respectively.
As such, one needs to focus only on the sub-DRTs pertinent to the AOS. Specifically, the nodes at level $n_e$ can only be
$(0,0)=:\mathrm{V}$, $(0,1)=:\mathrm{D}$, $(0,2)=:\mathrm{T}$, and $(1,0)=:\mathrm{S}$ for zero, one, two same-spin, and two opposite-spin electrons in the EOS,
respectively. Likewise, the nodes at level $n_e+n_a$ are
$(a,b)=(\frac{N_a}{2}-S,2S)=:\bar{\mathrm{V}}$, $(a,b+1)=:\bar{\mathrm{D}}$,
$(a,b+2)=:\bar{\mathrm{T}}$, and $(a+1,b)=:\bar{\mathrm{S}}$ for the excitations of zero, one,
two same-spin, and two opposite-spin electrons from the DOS to the AOS, respectively.
However, these are complete only when the total spin $S$ is zero. When $S=1/2$, there exists
an additional node $(a+1,b-1)=:\bar{\mathrm{D}}_{-1/2}$ for the excitation of a beta electron
from the DOS to the AOS. In this case, $\bar{\mathrm{D}}$ ($=\bar{\mathrm{D}}_{1/2}$)
refers to the excitation of a single alpha electron from the DOS to the AOS.
When $S\ge 1$, still one more node $(a+2,b-2)=:\bar{\mathrm{T}}_{-1}$ is required for
the excitation of two beta electrons from the DOS to the AOS.
In this case, there exist 6 nodes, with $\bar{\mathrm{T}}$ ($=\bar{\mathrm{T}}_{+1}$)
representing the excitation of two alpha electrons from the DOS to the AOS.
Having determined such nodes for a given spin $S$, the sub-DRTs ($\bar{\mathrm{X}}\mathrm{Y}$) (X, Y = V, D, T, S)
can readily be established by following the \textbf{abc-rule} outlined in Sec. S1 in the Supporting Information.
It follows that a sub-DRT ($\bar{\mathrm{X}}\mathrm{Y}$) is composed of nodes and arcs.
The nodes are enumerated sequentially from the head $\bar{\mathrm{X}}$ (No. $N^{\bar{\mathrm{X}}}_{\bar{\mathrm{X}}\mathrm{Y}}$) to the tail $\mathrm{Y}$ (No. $N^{\mathrm{Y}}_{\bar{\mathrm{X}}\mathrm{Y}}$).
A node $(a_w, b_w)_J$ at level $w$ may be connected downward up to four nodes $(a_{w-1},b_{w-1})_{J_{d\downarrow}}$ at level $w-1$ via
the arcs $J \rightarrow J_{d\downarrow}$ ($d_{\downarrow} \in [0,3]$). It can also be connected upward to
several nodes $(a_{w+1},b_{w+1})_{J_{d\uparrow}}$ at level $w+1$ via the possible arcs $J_{d\uparrow} \rightarrow J$
($d_{\uparrow} \in [0,3]$). Every node (arc) $K$ can be associated with a weight $X_K$ ($Y_{K_d}$), so as to
identify the in total $\dim(\bar{\mathrm{X}}\mathrm{Y})$ possible step vectors $\{(\mathbf{d}_a)_{\mu}=(d_{n_e+1}\cdots d_{n_e+n_a})_{\mu}\}$ of the sub-DRT (see Sec. S1 in the Supporting Information).
Once this is done, the step vectors $\{\mathbf{d}_{\mu}=(\mathbf{d}_e\mathbf{d}_a\mathbf{d}_h)_{\mu}\}$ representing
the reference and excited CSFs can readily be established. Here,
$(\mathbf{d}_e)_\mu=(d_1\cdots d_{n_e})_{\mu}$ and
$(\mathbf{d}_h)_\mu=(d_{n_e+n_a+1}\cdots d_{n_e+n_a+n_h})_\mu$ are the partial step vectors for the EOS and  DOS, respectively.
Note in passing that a bra vector representing $\langle\Phi_\mu|$ is to be written as $\langle \mathbf{d}_{\mu}|=\langle (\mathbf{d}_h\mathbf{d}_a\mathbf{d}_e)_{\mu}|$.

\begin{figure}
\centering
\includegraphics[width=0.7\textwidth]{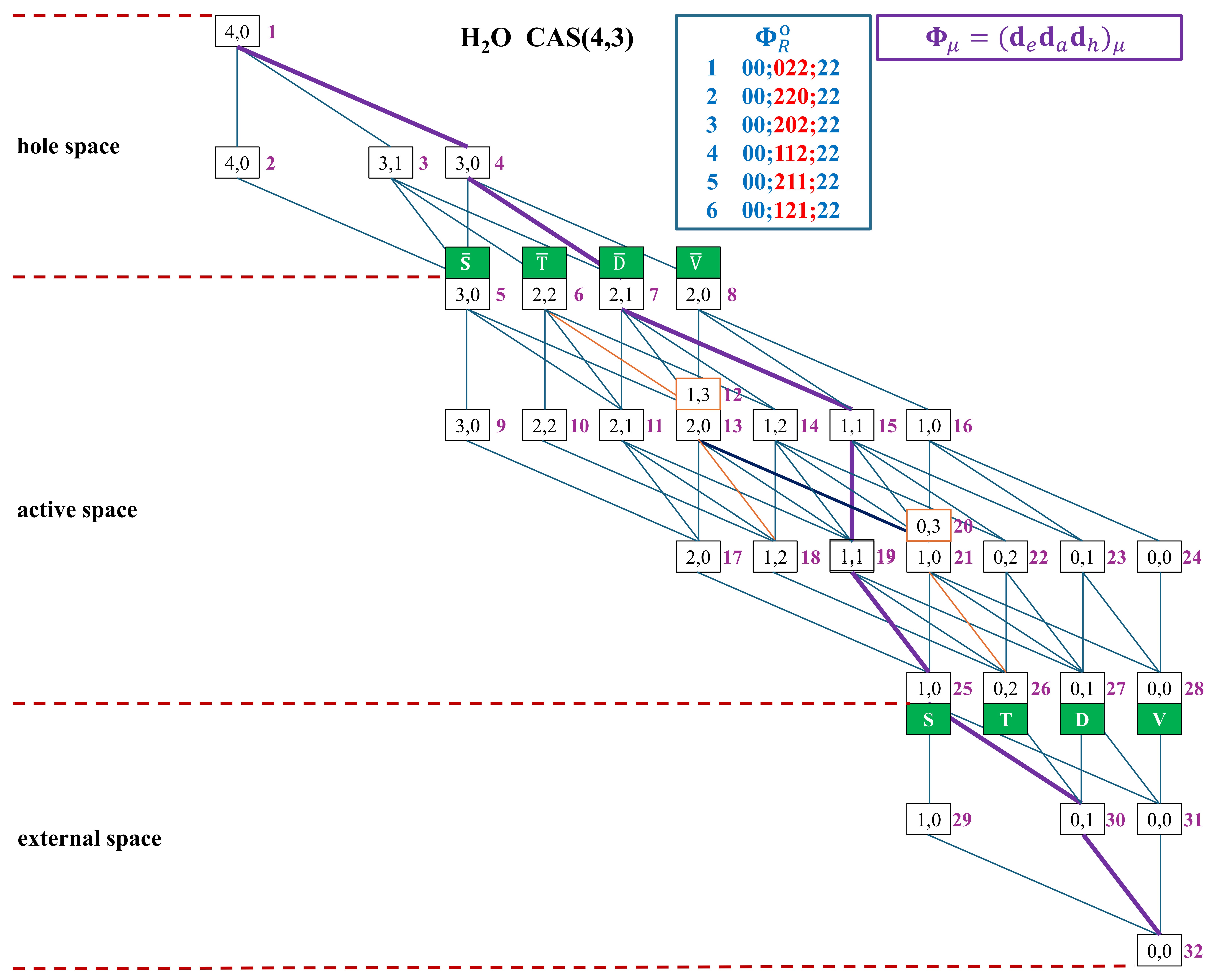}
\caption{DRT of frozen-core H$_2$O 
[$N=8$, $S=0$, $n=7$ (2 hole, 3 active, and 2 external orbitals)].
Inside and outside a box at level $r$ are node $(a_r,b_r)$ and its index $J$, respectively.
The blue solid line connecting two nodes at two adjacent levels is an arc.
Nodes at the same site (e.g., 12 and 13; 20 and 21) are distinguished by different colours.
The thick black arc indicates that node 12 (13) is connected to node 20 (21).
The graphs of the hole and external spaces are symmetric due to hole-particle correspondence.
In total 259 reference and excited CSFs are encoded by 32 nodes and their connections.
The thick purple arcs from the tail to the head denote the excited CSF
$\Phi_{188}=(\mathbf{d}_e\mathbf{d}_a\mathbf{d}_h)_{188}=\mathrm{1210323}$.}
\label{DRT-MRCISD}
\end{figure}

As illustrated in Fig. \ref{DRT-MRCISD} for constructing the CAS(4,3)-based uncontracted MRCISD space of H$_2$O,
the above procedure for constructing the excited CSFs from single and double excitations of a CAS $P$
is truly simple and efficient (see \textbf{Algorithm S1} in the Supporting Information).
However, the situation is different for a reduced space $\tilde{P}$ to be considered here.
The primary reason lies in that one need not pay attention to the reference oCFGs $\{\Phi_R^{\mathrm{o}}\}$ in the case of CAS
(for all of them are taken into account automatically), whereas in the case of $\tilde{P}$, not all reference oCFGs of the CAS are present.
A naive procedure is to delete those unwanted CSFs from the CAS-DRT. However, this demands a CSF-driven algorithm.
The merit of GUGA for computing basic coupling coefficients in terms of DRT loops is hence lost completely. It is shown here that
the reduced sub-DRTs can be constructed directly by imposing suitable constraints.

To construct a reduced sub-DRT ($\bar{\mathrm{X}}\mathrm{Y}$), the very first information to be employed
is the available number $M_{w,J}$ of electrons associated to each node $(a_w, b_w)_J$,
which is just the cumulative number of electrons starting from node Y (= V, D, T, S) at level $n_e$
 through each level upward to node $(a_w, b_w)_J$, and can simply be calculated as $2a_w+b_w$.
It is clear that $M_{n_e,N^{\mathrm{Y}}_{\bar{\mathrm{X}}\mathrm{Y}}}=0, 1, 2, 2$ for V, D, T, S
and $M_{n_e+n_a,N^{\bar{\mathrm{X}}}_{\bar{\mathrm{X}}\mathrm{Y}}}=N_a, N_a+1, N_a+2, N_a+2, N_a+1, N_a+2$ for $\bar{\mathrm{V}}$, $\bar{\mathrm{D}}$, $\bar{\mathrm{T}}$, $\bar{\mathrm{S}}$, $\bar{\mathrm{D}}_{-1/2}$, $\bar{\mathrm{T}}_{-1}$, respectively. Going down from level $w$ ($\le n_e+n_a$) to level $w-1$ ($\ge n_e+1$),
$M_{w,J}$ can be reduced by $N_{w,J}\in[0,2]$, leading to $M_{w-1,J_{d\downarrow}}$. Here, $N_{w,J}$
is just the occupation number of several excited oCFGs $\{\Phi_q^{\mathrm{o}}\}$ sharing node $(a_w, b_w)_J$
at level $w$. However, not all the three values of $N_{w,J}$ are allowed, since
the differential occupation $N_{w,J}-N_{w,R}^{\mathrm{o}}$ between the excited $\{\Phi_q^{\mathrm{o}}\}$ and a reference oCFG $\Phi_R^{\mathrm{o}}$ cannot be negative.
Moreover, when
going down from node $\bar{\mathrm{X}}$ through each level to node $(a_{w-1}, b_{w-1})_{J_{d\downarrow}}$,
the cumulative excitation number $\Delta N^{w-1}_{R,J_{d\downarrow}}$ relative to the reference oCFG $\Phi_R^{\mathrm{o}}$
is also limited by the total excitation number
$N^{ex}_{\bar{\mathrm{X}}\mathrm{Y}}$ of sub-DRT ($\bar{\mathrm{X}}\mathrm{Y}$), as documented in Table \ref{sub-DRT}.
$\Delta N^{w-1}_{R,J_{d\downarrow}}$ can be calculated recursively in a top-down manner as
\begin{equation}\label{Tcal}
\Delta N^{w-1}_{R,J_{d\downarrow}}=
\begin{cases}
\Delta N^{w}_{R,J}+(N_{w,J}-N_{w,R}^{\mathrm{o}}) & \text{if $N_{w,J}> N_{w,R}^{\mathrm{o}}$,} \\
\Delta N^{w}_{R,J} & \text{if $N_{w,J}\leq N_{w,R}^{\mathrm{o}}$,}
\end{cases}
\end{equation}
where $w\in[n_e+2, n_e+n_a]$ and $\Delta N^{n_e+n_a}_{R,1}=0$.
When $\Delta N^{w-1}_{R,J_{d\downarrow}}$ is greater than the maximum of $N^{ex}_{\bar{\mathrm{X}}\mathrm{Y}}$, the set of excited oCFGs $\{\Phi_q^{\mathrm{o}}\}$
passing through node $(a_{w-1}, b_{w-1})_{J_{d\downarrow}}$  will be independent of the reference oCFG $\Phi_R^{\mathrm{o}}$. The node $(a_{w-1}, b_{w-1})_{J_{d\downarrow}}$
will be invalid when the minimal value, $\min_{R} \Delta N^{w-1}_{R,J_{d\downarrow}}$,
of the whole set $\mathbf{V}_{w-1}=\{\Delta N^{w-1}_{R,J_{d\downarrow}}; w\in[n_e+2, n_e+n_a]\}_{R=1}^{N_R^{\mathrm{o}}}$
is greater than $N^{ex}_{\bar{\mathrm{X}}\mathrm{Y}}$, since in this case
all excited oCFGs would have excitation numbers larger than the allowed $N^{ex}_{\bar{\mathrm{X}}\mathrm{Y}}$.
Once the valid $N_{w,J}$ are determined, the allowed step values $d_{w}\in [0,3]$ and hence the arcs $J\rightarrow J_{d\downarrow}$
can be fixed following the \textbf{abc-rule} in the Supporting Information.

However, the above procedure still ends up with many redundant intermediate nodes. To see this,
 we first define the remaining number $M^{re}_{w-1,R}$ of active electrons of the reference oCFG $\Phi_R^{\mathrm{o}}$ at level $w-1$ as
\begin{eqnarray}
M^{re}_{w-1,R}=N_a-\sum_{u=w}^{n_e+n_a} N_{u,R}^{\mathrm{o}}.
\end{eqnarray}
It is obvious that $M_{w-1,J_{d\downarrow}}$ is bounded by $M^{re}_{w-1,R}+M^{ex}_{\bar{\mathrm{X}}\mathrm{Y}}$ with
$M^{ex}_{\bar{\mathrm{X}}\mathrm{Y}}=2$.
That $M_{w-1,J_{d\downarrow}} > M^{re}_{w-1,R} + M^{ex}_{\bar{\mathrm{X}}\mathrm{Y}}$ means that
more than two electrons have been excited from the internal orbitals $p \in [w, n]$ of $\Phi_R^{\mathrm{o}}$,
thereby rendering $\Phi_R^{\mathrm{o}}$ deadwood for the intermediate node $(a_{w-1}, b_{w-1})_{J_{d\downarrow}}$.
The survival $\Delta N^{w-1}_{R,J_{d\downarrow}}$ can be collected
into the effective set $\textbf{V}_{w-1}^{eff}=\{\Delta N^{w-1}_{R,J_{d\downarrow}}; w\in[n_e+2, n_e+n_a]\}$.
If the following condition
\begin{eqnarray}\label{Tcheck}
\textbf{V}_{w-1}^{eff}=\emptyset \mbox{ or } \min_R(\textbf{V}_{w-1}^{eff})> N^{ex}_{\bar{\mathrm{X}}\mathrm{Y}}
\end{eqnarray}
is satisfied, $\{\Phi_q^{\mathrm{o}}\}$ with $N_{w,J}$ will not be generated from single and double excitations from $\{\Phi_R^{\mathrm{o}}\}_{R=1}^{N_R^{\mathrm{o}}}$. The intermediate nodes $(a_{w-1}, b_{w-1})_{J_{d\downarrow}}$ and hence
the corresponding arcs $J \rightarrow J_{d\downarrow}$ should be deleted.
Moreover, if $w=n_e+2$, $(a_{w-1}, b_{w-1})_{J_{d\downarrow}}$ must be one of the nodes
$(a_{n_e+1},b_{n_e+1})_d$ ($d\in[0,3]$) that match Y (see Table \ref{Yabup}); otherwise, delete $(a_{w-1}, b_{w-1})_{J_{d\downarrow}}$.

\begin{figure}[htbp]
	\centering
	\begin{subfigure}
		\centering
        \includegraphics[width=0.49\textwidth]{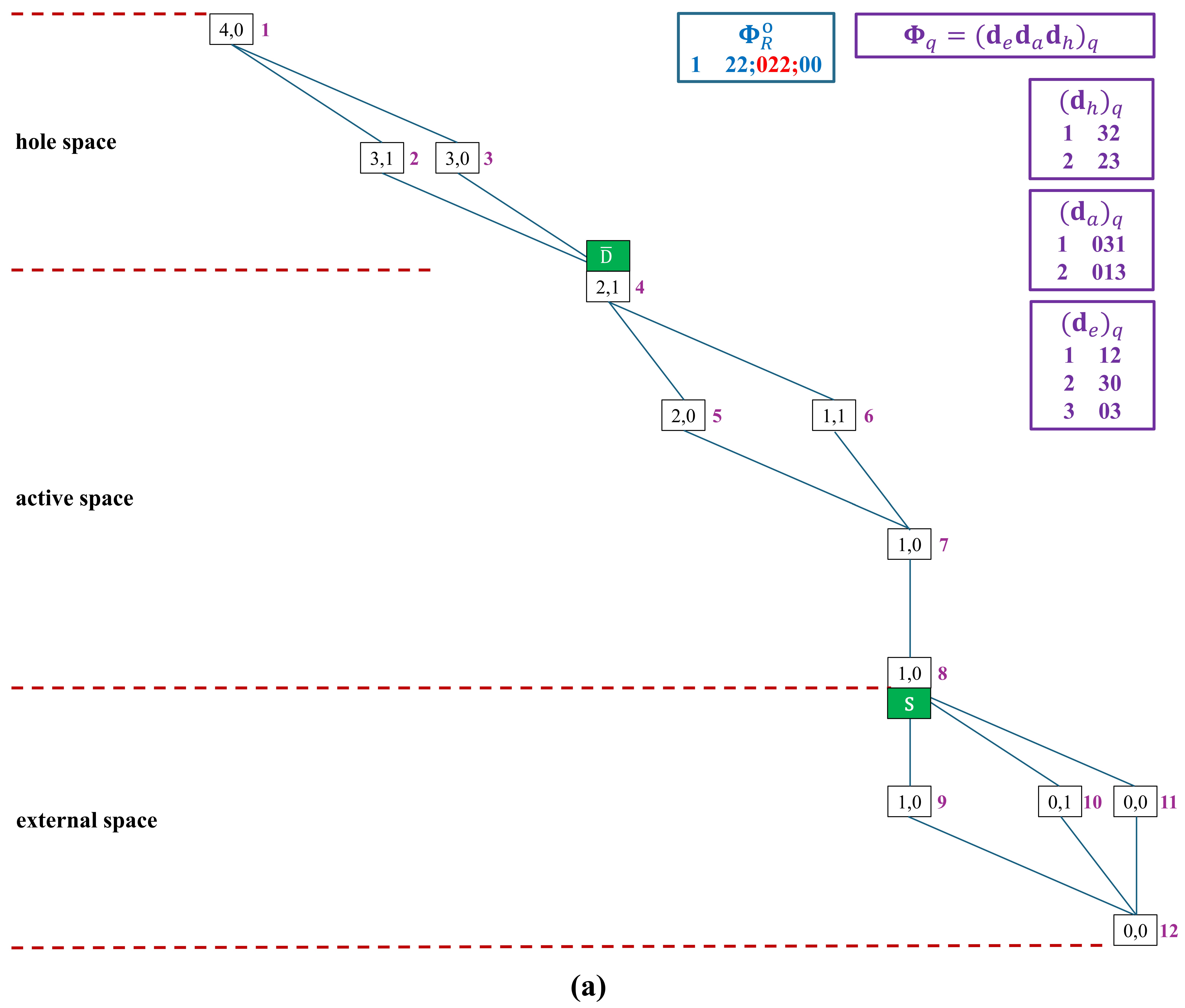}
	\end{subfigure}
    \centering
	\begin{subfigure}
		\centering
        \includegraphics[width=0.49\textwidth]{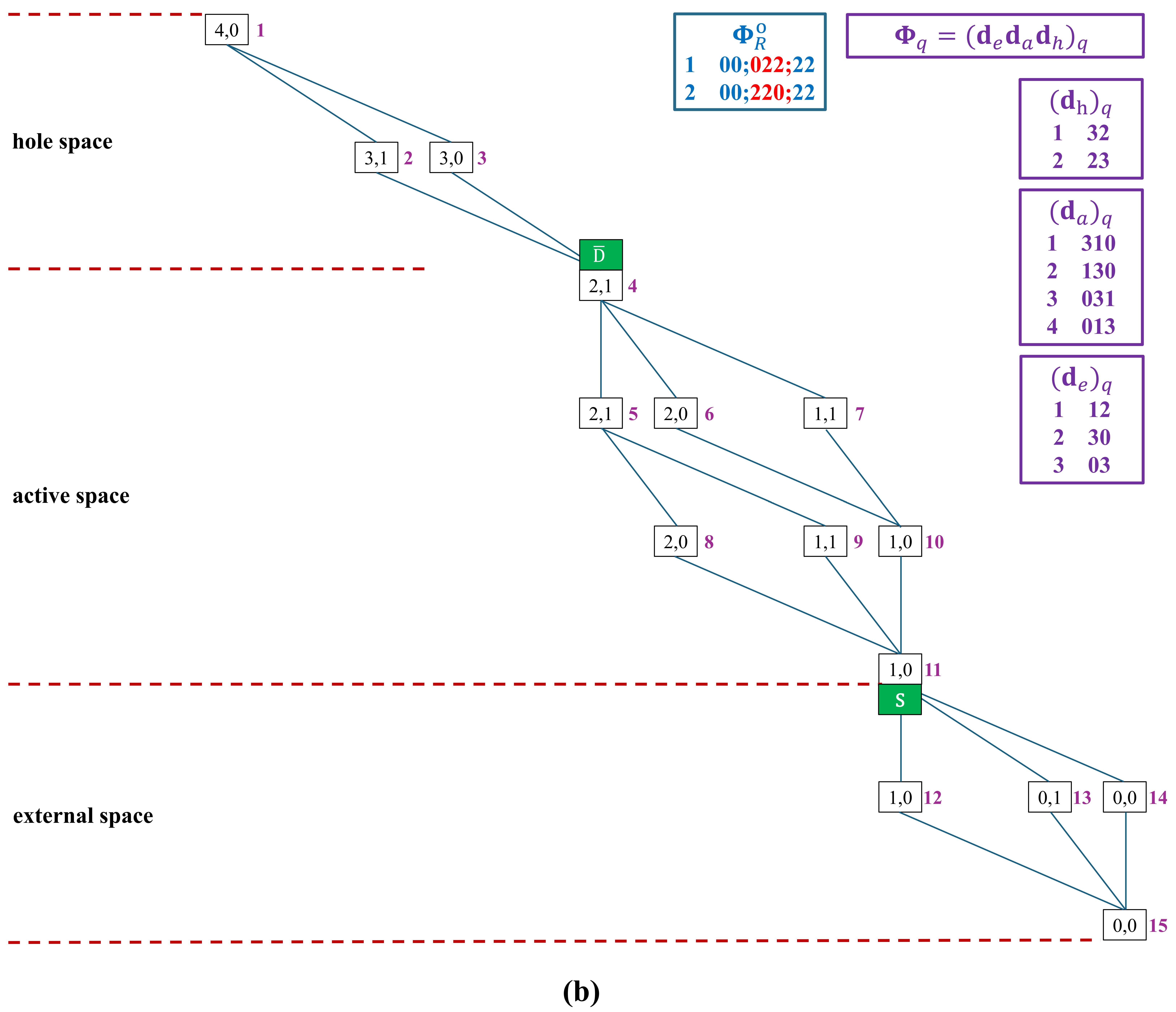}
	\end{subfigure}

    \caption{Reduced sub-DRT ($\bar{\mathrm{D}}\mathrm{S}$) of frozen-core H$_2$O 
[$N=8$, $S=0$, $n=7$ (2 hole, 3 active, and 2 external orbitals)].
(a) one reference oCFG ($\Phi_R^{\mathrm{o}}$); (b) two reference oCFGs.
Inside and outside a box at level $r$ are node $(a_r,b_r)$ and its index $J$, respectively.
The blue solid line connecting two nodes at two adjacent levels is an arc.
$\{\Phi_q\}$ are excited CSFs.}
    \label{DS-MRCISD}
\end{figure}

It should also be noted that, while the nodes at different levels are inherently distinct, the nodes at the same level
must be made unique by double checking with the existing ones. To achieve this, the multidimensional array $(a_{w-1},b_{w-1},\textbf{V}_{w-1}^{eff})_{J_{d\downarrow}}$
has to be stored. A node $J_{d\downarrow}$ is considered new only if its metadata $(a_{w-1},b_{w-1},\textbf{V}_{w-1}^{eff})$ differs from all the existing ones at the same level $w-1$.
This double checking process is the rate determining step for constructing the reduced sub-DRTs $(\bar{\mathrm{X}}\mathrm{Y})$.
After this, all intermediate nodes of $(\bar{\mathrm{X}}\mathrm{Y})$ have been generated and $\textbf{V}_{w-1}^{eff}$ is no longer needed.
Starting from Y to $\bar{\mathrm{X}}$, the weights of the nodes can readily be calculated according Eq. (S1).
The nodes with the same weight and downward connecting nodes $J_{d\downarrow}$ are actually identical and should hence be merged together.
The survival nodes and corresponding arcs finally form ($\bar{\mathrm{X}}\mathrm{Y}$). As an illustration,
the detailed derivations of the reduced sub-DRT ($\bar{\mathrm{D}}\mathrm{S}$) for H$_2$O  with one and two oCFGs are documented in Tables \ref{DS1} and \ref{DS2},
respectively. The results are further plotted in Fig. \ref{DS-MRCISD}.

The above procedure for constructing the reduced sub-DRTs can be implemented by \textbf{Algorithm 1}.
Compared with \textbf{Algorithm S1} in the Supporting Information for constructing the full sub-DRTs,
only lines 11--12 and 14--18 are new, which perform validation of
nodes $(a_{w-1}, b_{w-1})_{J_{d\downarrow}}$ [cf. Eqs. \eqref{Tcal} and \eqref{Tcheck}] and double checking, respectively.
Since the number of nodes is much less than that of CSFs, the present `reduced sub-DRT' scheme
should be more efficient than those working directly with randomly selected CSFs.

Noticeably, ($\bar{\mathrm{D}}\mathrm{V}$) and ($\bar{\mathrm{V}}\mathrm{D}$) are the largest sub-DRTs due to the involvement
of three active orbitals (see Table \ref{sub-DRT}). They can be reduced by setting
$M^{ex}_{\bar{\mathrm{D}}\mathrm{V}} = M^{ex}_{\bar{\mathrm{V}}\mathrm{D}} = 1$ in lieu of $M^{ex}_{\bar{\mathrm{D}}\mathrm{V}} = M^{ex}_{\bar{\mathrm{V}}\mathrm{D}} = 2$.
This amounts to screening out some three-active-orbital double (TAOD) excitations that
describe the interactions between a hole (in the case of ($\bar{\mathrm{D}}\mathrm{V}$))
or particle (in the case of ($\bar{\mathrm{V}}\mathrm{D}$)) with an active electron, mediated
by the scattering of the other active electrons. It will be shown in Sec. \ref{Truncation} that
this is a very good approximation (denoted as DVD). However, it is dependent on the ordering of the active orbitals
and may hence be used safely only for calculations of vertical excitations.

\vspace{1cm}
\begin{threeparttable}
\tabcolsep=10pt
\scriptsize
\begin{tabular}{ll}
  \hline
  \textbf{Algorithm 1}. Construction of reduced sub-DRTs\\
  \hline
  01:\quad  \ \textbf{for} $\bar{\mathrm{X}}\in$ nodes at level $n_e+n_a$ \ \textbf{do} \\
  02:\quad  \ \ \textbf{for} $\mathrm{Y}\in$ nodes at level $n_e$ \ \textbf{do} \\
  03:\quad  \ \ \ \ \textbf{for} $w \in [n_e+n_a,n_e+2]$; \ $w--$; \ \textbf{do}\\
  04:\quad  \ \ \ \ \ \ search for node $J_{d\downarrow}$ at level $w-1$ from node $J$ at level $w$ following the \textbf{abc-rule} \\
  05:\quad  \ \ \ \ \ \ \textbf{if} $w=n_e+2$ \ \textbf{then} \\
  06:\quad  \ \ \ \ \ \ \ \ \textbf{if} $(a_{n_e+1},b_{n_e+1})$ of $J_{d\downarrow}=(a_{n_e+1},b_{n_e+1})_d$ ($d\in[0,3]$) for Y in Table \ref{Yabup} \ \textbf{then} \\
  07:\quad  \ \ \ \ \ \ \ \ \ \ connect $J_{d\downarrow}$ with Y \\
  08:\quad  \ \ \ \ \ \ \ \ \textbf{else} \\
  09:\quad  \ \ \ \ \ \ \ \ \ \ remove $J_{d\downarrow}$  \\
  10:\quad  \ \ \ \ \ \ \textbf{else} \\
  11:\quad  \ \ \ \ \ \ \ \ \textbf{if} Eq. (\ref{Tcheck}) is satisfied for the set $\textbf{V}_w^{eff}$ \ \textbf{then} \\
  12:\quad  \ \ \ \ \ \ \ \ \ \ remove $J_{d\downarrow}$  \\
  13:\quad  \ \ \ \ \ \ \ \ \textbf{else} \\
  14:\quad  \ \ \ \ \ \ \ \ \ \ \textbf{for} $J^\prime < J_{d\downarrow}$ at level $w-1$; \ $J^\prime++$; \ \textbf{do} \\
  15:\quad  \ \ \ \ \ \ \ \ \ \ \ \ \textbf{if} $(a_{w-1},b_{w-1},\textbf{V}_{w-1}^e)$ of $J$ is not equal to $(a_{w-1},b_{w-1},\textbf{V}_{w-1}^{eff})$ of ${J^\prime}$ \ \textbf{then}\\
  16:\quad  \ \ \ \ \ \ \ \ \ \ \ \ \ \ $J_{d\downarrow}$ is new and connect $J$ with $J_{d\downarrow}$ \\
  17:\quad  \ \ \ \ \ \ \ \ \ \ \ \ \textbf{else}  \\
  18:\quad  \ \ \ \ \ \ \ \ \ \ \ \ \ \ remove $J_{d\downarrow}$ and connect $J$ with $J^\prime$\\
  19-34: \ same as Lines 14--29 of \textbf{Algorithm S1} in the Supporting Information for removing redundant nodes. \\
  35:\quad  \ \ \ \ \textbf{for} $w \in [n_e+n_a,n_e+1]$; \ $w--$; \ \textbf{do} \\
  36:\quad  \ \ \ \ \ \ \textbf{for} $J$ on $w$-th level; \ $J++$; \ \textbf{do} \\
  37:\quad  \ \ \ \ \ \ \ \ calculate $X_J$ and $Y_{J_d}$ of the survival $J$ and determine the indices of $(\mathbf{d}_a)_\mu$ \\
  \hline
\end{tabular}
\end{threeparttable}
\vspace{1cm}

\begin{threeparttable}
  \caption{Possible upper nodes $(a_{n_e+1},b_{n_e+1})_d$ of node Y at level $n_e$}
  \tabcolsep=20pt
  \scriptsize
  \begin{tabular}[h]{@{}cccccccccccccccccc@{}}
   \hline\hline
   Y & $(a_{n_e+1},b_{n_e+1})_3$ & $(a_{n_e+1},b_{n_e+1})_2$ & $(a_{n_e+1},b_{n_e+1})_1$ & $(a_{n_e+1},b_{n_e+1})_0$ \\
   \hline
   V & (1,0) &       & (0,1) & (0,0) \\
   D & (1,1) & (1,0) & (0,2) & (0,1) \\
   T & (1,2) & (1,1) & (0,3) & (0,2) \\
   S & (2,0) &       & (1,1) & (1,0) \\
   \hline \hline
  \end{tabular}\label{Yabup}
\end{threeparttable}
\vspace{1cm}

\begin{threeparttable}
  \caption{Derivation of sub-DRT ($\bar{\mathrm{D}}\mathrm{S}$) for the orbital configuration $\Phi^{\mathrm{o}}_1=00;022;22$ of frozen-core H$_2$O 
[$N=8$, $S=0$, $n=7$ (2 hole, 3 active, and 2 external orbitals)]}
  \tabcolsep=10pt
  \scriptsize
  \begin{tabular}[h]{@{}cccccccccccccccccc@{}}
   \hline\hline
$J$ & $w$ & $N_{w,J}$ & $N_{w,1}^{\mathrm{o}}$ & $\Delta N^{w-1}_{1,J_{d\downarrow}}$ & $M_{w-1,J_{d\downarrow}}$ & $M^{re}_{w-1,1}$ & $J_{d\downarrow}$ & $(a_{w-1},b_{w-1})$ & Inode\tnote{a} & Fnode\tnote{b}\\
   \hline
4	& 5	& 0	& 2	& 0	& 5	& 2	&5	&(2,1)	& \textcolor[rgb]{0.00,0.07,1.00}{no\tnote{c}}	& no\\
4	& 5	& 1	& 2	& 0	& 4	& 2	&6	&(2,0)	& yes	& yes\\
4	& 5	& 1	& 2	& 0	& 4	& 2	&7	&(1,2)	& yes	& \textcolor[rgb]{1.00,0.00,0.00}{no\tnote{e}}\\
4	& 5	& 2	& 2	& 0	& 3	& 2	&8	&(1,1)	& yes	& yes\\
6	& 4	& 0	& 2	& 0	& 4	& 0	&9	&(2,0)	& \textcolor[rgb]{0.00,0.07,1.00}{no\tnote{c}}	& no\\
6	& 4	& 1	& 2	& 0	& 3	& 0	&10	&(1,1)	& \textcolor[rgb]{0.00,0.07,1.00}{no\tnote{c}}	& no\\
6	& 4	& 2	& 2	& 0	& 2	& 0	&11	&(1,0)	& yes	& yes\\
7	& 4	& 0	& 2	& 0	& 4	& 0	&12	&(1,2)	& \textcolor[rgb]{0.50,0.00,0.50}{no\tnote{d}}	& no\\
7	& 4	& 1	& 2	& 0	& 3	& 0	&13	&(1,1)	& \textcolor[rgb]{0.00,0.07,1.00}{no\tnote{c}}	& no\\
7	& 4	& 1	& 2	& 0	& 3	& 0	&14	&(0,3)	& \textcolor[rgb]{0.50,0.00,0.50}{no\tnote{d}}	& no\\
7	& 4	& 2	& 2	& 0	& 2	& 0	&15	&(0,2)	& \textcolor[rgb]{0.50,0.00,0.50}{no\tnote{d}}	& no\\
8	& 4	& 0	& 2	& 0	& 3	& 0	&16	&(1,1)	& \textcolor[rgb]{0.00,0.07,1.00}{no\tnote{c}}	& no\\
8	& 4	& 1	& 2	& 0	& 2	& 0	&17	&(1,0)	& yes	& \textcolor[rgb]{1.00,0.00,0.00}{no\tnote{e}}\\
8	& 4	& 1	& 2	& 0	& 2	& 0	&18	&(0,2)	& \textcolor[rgb]{0.50,0.00,0.50}{no\tnote{d}}	& no\\
8	& 4	& 2	& 2	& 0	& 1	& 0	&19	&(0,1)	& \textcolor[rgb]{0.50,0.00,0.50}{no\tnote{d}}	& no\\
11	& 3	& 0	& 0	& 0	& 2	& 0	&20	&(1,0)	& yes	& yes\\
17	& 3	& 0	& 0	& 0	& 2	& 0	&21	&(1,0)	& yes	& \textcolor[rgb]{0.00,0.46,0.29}{no\tnote{f}}\\
   \hline \hline
  \end{tabular}\label{DS1}
  \begin{tablenotes}
  \item[] $^\mathrm{a}$ Intermediate node; $^\mathrm{b}$ Final node; $^\mathrm{c}$ $\textbf{V}_{w-1}^{eff}=\emptyset$ in Eq. (\ref{Tcheck});
  $^\mathrm{d}$ $(a_{n_e+1},b_{n_e+1})$ of $J_{d\downarrow}\neq(a_{n_e+1},b_{n_e+1})_d$ ($d\in[0,3]$) for Y = S in Table \ref{Yabup};
  $^\mathrm{e}$ disconnected; $^\mathrm{f}$ merged.
  \end{tablenotes}
\end{threeparttable}
\vspace{1cm}

\begin{threeparttable}
  \caption{Derivation of sub-DRT ($\bar{\mathrm{D}}\mathrm{S}$) for the orbital configurations $\Phi^{\mathrm{o}}_1=00;022;22$ and $\Phi^{\mathrm{o}}_2=00;220;22$ of frozen-core H$_2$O  [$N=8$, $S=0$, $n=7$ (2 hole, 3 active, and 2 external orbitals)]}
  \tabcolsep=6pt
  \scriptsize
  \begin{tabular}[h]{@{}cccccccccccccccccccccccccccc@{}}
   \hline\hline
$J$ & $w$ & $N_{w,J}$ & $N_{w,1}^{\mathrm{o}}$ & $N_{w,2}^{\mathrm{o}}$ & $\Delta N^{w-1}_{1,J_{d\downarrow}}$ & $\Delta N^{w-1}_{2,J_{d\downarrow}}$ & $M_{w-1,J_{d\downarrow}}$ & $M^{re}_{w-1,1}$ & $M^{re}_{w-1,2}$ & $J_{d\downarrow}$ & $(a_{w-1},b_{w-1})$ & Inode\tnote{a} & Fnode\tnote{b}\\
   \hline
4	& 5	& 0	& 2	& 0	& 0	& 0	& 5	& 2	& 4	&5	&(2,1)	& yes	& yes\\
4	& 5	& 1	& 2	& 0	& 0	& 1	& 4	& 2	& 4	&6	&(2,0)	& yes	& yes\\
4	& 5	& 1	& 2	& 0	& 0	& 1	& 4	& 2	& 4	&7	&(1,2)	& yes	& \textcolor[rgb]{1.00,0.00,0.00}{no\tnote{e}}\\
4	& 5	& 2	& 2	& 0	& 0	& 2	& 3	& 2	& 4	&8	&(1,1)	& yes	& yes\\
5	& 4	& 0	& 2	& 2	& 0	& 0	& 5	& 0	& 2	&9	&(2,1)	& \textcolor[rgb]{0.50,0.00,0.50}{no\tnote{c}}	& no\\
5	& 4	& 1	& 2	& 2	& 0	& 0	& 4	& 0	& 2	&10	&(2,0)	& yes	& yes\\
5	& 4	& 1	& 2	& 2	& 0	& 0	& 4	& 0	& 2	&11	&(1,2)	& \textcolor[rgb]{0.50,0.00,0.50}{no\tnote{c}}	& no\\
5	& 4	& 2	& 2	& 2	& 0	& 0	& 3	& 0	& 2	&12	&(1,1)	& yes	& yes\\
6	& 4	& 0	& 2	& 2	& 0	& 1	& 4	& 0	& 2	&13	&(2,0)	& yes	& \textcolor[rgb]{1.00,0.00,0.00}{no\tnote{e}}\\
6	& 4	& 1	& 2	& 2	& 0	& 1	& 3	& 0	& 2	&14	&(1,1)	& yes	& \textcolor[rgb]{1.00,0.00,0.00}{no\tnote{e}}\\
6	& 4	& 2	& 2	& 2	& 0	& 1	& 2	& 0	& 2	&15	&(1,0)	& yes	& yes\\
7	& 4	& 0	& 2	& 2	& 0	& 1	& 4	& 0	& 2	&16	&(1,2)	& \textcolor[rgb]{0.50,0.00,0.50}{no\tnote{c}}	& no\\
7	& 4	& 1	& 2	& 2	& 0	& 1	& 3	& 0	& 2	&17	&(1,1)	& yes	& \textcolor[rgb]{1.00,0.00,0.00}{no\tnote{e}}\\
7	& 4	& 1	& 2	& 2	& 0	& 1	& 3	& 0	& 2	&18	&(0,3)	& \textcolor[rgb]{0.50,0.00,0.50}{no\tnote{c}}	& no\\
7	& 4	& 2	& 2	& 2	& 0	& 1	& 2	& 0	& 2	&19	&(0,2)	& \textcolor[rgb]{0.50,0.00,0.50}{no\tnote{c}}	& no\\
8	& 4	& 0	& 2	& 2	& 0	& 2	& 3	& 0	& 2	&20	&(1,1)	& yes	& \textcolor[rgb]{1.00,0.00,0.00}{no\tnote{e}}\\
8	& 4	& 1	& 2	& 2	& 0	& 2	& 2	& 0	& 2	&21	&(1,0)	& yes	& \textcolor[rgb]{0.00,0.46,0.29}{no\tnote{f}}\\
8	& 4	& 1	& 2	& 2	& 0	& 2	& 2	& 0	& 2	&22	&(0,2)	& \textcolor[rgb]{0.50,0.00,0.50}{no\tnote{c}}	& no\\
8	& 4	& 2	& 2	& 2	& 0	& 2	& 1	& 0	& 2	&23	&(0,1)	& \textcolor[rgb]{0.50,0.00,0.50}{no\tnote{c}}	& no\\
10	& 3	& 2	& 0	& 2	& 2	& 0	& 2	& 0	& 0	&24	&(1,0)	& yes	& yes\\
12	& 3	& 1	& 0	& 2	& 1	& 0	& 2	& 0	& 0	&25	&(1,0)	& yes	& \textcolor[rgb]{0.00,0.46,0.29}{no\tnote{f}}\\
13	& 3	& 2	& 0	& 2	& 2	& 1	& 2	& 0	& 0	&26	&(1,0)	& \textcolor[rgb]{0.00,0.07,1.00}{no\tnote{d}}	& no\\
14	& 3	& 1	& 0	& 2	& 1	& 1	& 2	& 0	& 0	&27	&(1,0)	& \textcolor[rgb]{0.00,0.07,1.00}{no\tnote{d}}	& no\\
15	& 3	& 0	& 0	& 2	& 0	& 1	& 2	& 0	& 0	&28	&(1,0)	& yes	& \textcolor[rgb]{0.00,0.46,0.29}{no\tnote{f}}\\
17	& 3	& 1	& 0	& 2	& 1	& 1	& 2	& 0	& 0	&29	&(1,0)	& \textcolor[rgb]{0.00,0.07,1.00}{no\tnote{d}}	& no\\
20	& 3	& 1	& 0	& 2	& 1	& 2	& 2	& 0	& 0	&30	&(1,0)	&\textcolor[rgb]{0.00,0.07,1.00}{ no\tnote{d}}	& no\\
21	& 3	& 0	& 0	& 2	& 0	& 2	& 2	& 0	& 0	&31	&(1,0)	& yes	& \textcolor[rgb]{0.00,0.46,0.29}{no\tnote{f}}\\
   \hline \hline
  \end{tabular}\label{DS2}
  \begin{tablenotes}
  \item[] $^\mathrm{a}$ Intermediate node; $^\mathrm{b}$ Final node;
  $^\mathrm{c}$ $(a_{n_e+1},b_{n_e+1})$ of $J_{d\downarrow}\neq(a_{n_e+1},b_{n_e+1})_d$ ($d\in[0,3]$) for Y = S in Table \ref{Yabup};
  $^\mathrm{d}$ $\min_R(\textbf{V}_{w-1}^{eff})> N^{ex}_{\bar{\mathrm{D}}\mathrm{S}}=0$ in Eq. (\ref{Tcheck});
  $^\mathrm{e}$ disconnected; $^\mathrm{f}$ merged.
  \end{tablenotes}
\end{threeparttable}
\vspace{1cm}

\subsection{Matrix elements}\label{ic-MRCISD}
What has been discussed so far is how to generate the excited CSFs. However, even for the reduced reference space $\tilde{P}$,
the use of individual CSFs as perturbers is still prohibitively expensive. In contrast, the use of internally contracted configurations (ICC)
is more effective\cite{selectMRPT2}, for they are known to
span exactly the FOIS, even though the number of them is only related to the number of active orbitals
but independent of the number of reference CSFs. In the parlance of HPS-GUGA,
the ICCs are defined as\cite{WenicMRCI}
\begin{align}
\bar{\Phi}^{\bar{\mathrm{X}}\mathrm{Y}}_{Mk}&=\hat{E}_M|\Psi^{(0)}_k\rangle,\quad \hat{E}_M =\hat{E}_{pq}, \hat{e}_{pq,rs}\in\bar{\mathrm{X}}\mathrm{Y}\nonumber \\
&
= \sum_{q\in\bar{\mathrm{X}}\mathrm{Y}}|\Phi_q\rangle C_{qM}^k,\label{ic-definition}\\
C_{qM}^k&=\langle\Phi_q|\hat{E}_M|\Psi^{(0)}_k\rangle=\sum_{R\in\bar{\mathrm{V}}\mathrm{V}}
\langle\Phi_q|\hat{E}_M|\Phi_R\rangle \bar{C}_{{R}k}^{(0)},\label{coef-ic}
\end{align}
where the excitation operators have been defined in Table \ref{sub-DRT}.
The internal contraction coefficients $C_{qM}^k$ ($q\in\bar{\mathrm{X}}Y$) can be reexpressed as\cite{WenicMRCI}
\begin{align}\label{CM}
C_{qM}^k&=\sum_{R\in\bar{\mathrm{V}}\mathrm{V}}\langle (\mathbf{d}_h\mathbf{d}_a\mathbf{d}_e)_q|\hat{E}_M|
(\mathbf{d}_e\mathbf{d}_a\mathbf{d}_h)_R\rangle \bar{C}^{(0)}_{Rk} \\
&=ELS(\mathrm{YV}) \cdot HLS(\bar{\mathrm{X}}\bar{\mathrm{V}}) \cdot A^k_{q_aM},\label{CM-1}\\
A^k_{q_aM}&=\sum_{R_a\in(\bar{\mathrm{V}}\mathrm{V})}ALS(\bar{\mathrm{X}}\mathrm{Y},\bar{\mathrm{V}}\mathrm{V})_{q_aR_a} \bar{C}^{(0)}_{R_ak}.\label{aM}
\end{align}
Here, $q_a$ and $R_a$ are the lexical orders of step vectors $(\mathbf{d}_a)_{q}$ and $(\mathbf{d}_a)_{R}$, respectively
(see Sec. S1 in the Supporting Information).
$ELS(\mathrm{YV})$ (involving only external orbitals) and $HLS(\bar{\mathrm{X}}{\bar{\mathrm{V}}})$ (involving only
doubly occupied orbitals) are the external and hole loop shapes, respectively,
and can be pre-computed\cite{WenicMRCI}. As a result, the searching for loops needs to be performed only within the AOS,
which involves much fewer partial loops compared to the implementation without using the hole-particle correspondence.
Note that multiple $ALS(\bar{\mathrm{X}}\mathrm{Y},{\bar{\mathrm{V}}} \mathrm{V})_{q_a R_a}$ with identical active orbitals can be searched simultaneously
on the fly and contracted with  $\bar{C}^{(0)}_{Rk}$ in parallel with respect to partial loops (which are defined
by the ranges of the indices of the excitation operators, see Table 2 and Figures 2 and 3 in Ref. \citenum{iCIPT2}).

The ICCs $\{\bar{\Phi}^{\bar{\mathrm{X}}\mathrm{Y}}_{Mk}\}$ are not orthogonal to each other and are even linearly dependent. Therefore,
a canonical orthonormalization has to be carried out. Since both $HLS(\bar{\mathrm{X}}\bar{\mathrm{V}})$ and $ELS(\mathrm{YV})$ are just factors,
they can be absorbed in the orthonormalization procedure. As such, the contraction coefficients $C_{qM}^k$
can simply be set to $A^k_{q_a M}$ \eqref{aM}, which can be stored as a sparse matrix for each state $k$.
Those excited CSFs $\Phi_q$ with zero coefficients $C_{qM}^k$
belong to the non-interacting external space $\tilde{Q}_r$ and will hence be discarded.

As stated before, the $(\bar{\mathrm{D}}\mathrm{V})$ and $(\bar{\mathrm{V}}\mathrm{D})$ subspaces
are memory intensive, due to the fact that the numbers of coefficients $A^k_{q_a M}$ \eqref{aM} scale as
$n_a^3 \times \dim(\bar{\mathrm{V}}\mathrm{D})$ and $n_a^3 \times \dim(\bar{\mathrm{D}}\mathrm{V})$, respectively.
Apart from the previous DVD approximation, an integral-driven elimination of unimportant TAODs can further be introduced, viz.,
\begin{align}\label{Qmin}
\max_{p, R\in\bar{\mathrm{V}}\mathrm{V}}(|(pu|vw)\cdot \langle(\mathbf{d}_h\mathbf{d}_a\mathbf{d}_e)_q|\hat{e}_{pu,vw}|\Phi_R\rangle \bar{C}^{(0)}_{Rk}|)<Q_{\text{min}},
\end{align}
where $p$ refers to a doubly occupied and an external orbital for sub-DRTs $(\bar{\mathrm{D}}\mathrm{V})$ and $(\bar{\mathrm{V}}\mathrm{D})$, respectively,
and $Q_{\text{min}}$ is the prechosen threshold. Note that Eq. \eqref{Qmin} eliminates TAODs in batches instead of individually.
That is, given a partial step vector $\mathbf{d}_a$ [i.e., a specific triple $(u,vw)$ in the expression \eqref{Qmin}] and
a partial step vector $\mathbf{d}_e$/$\mathbf{d}_h$,
all TAODs $(\mathbf{d}_h\mathbf{d}_a\mathbf{d}_e)_q$ with different $(\mathbf{d}_h)_q$/$(\mathbf{d}_e)_q$
are pruned away if the condition (\ref{Qmin}) holds in the case of $(\bar{\mathrm{D}}\mathrm{V})$/$(\bar{\mathrm{V}}\mathrm{D})$.
It will be shown in Sec. \ref{Truncation} that such prescreening reduces the number of TAODs substantially
and hence also the number of ICCs. However, experimentations reveal that similar prescreenings cannot be applied
to the other CI subspaces.

The coupling coefficients between the ICCs can be calculated as
\begin{align}\label{CC-ic}
\langle\bar{\Phi}^{\bar{\mathrm{X}}\mathrm{Y}}_{Mk}|\hat{E}_M|\bar{\Phi}^{{\bar{\mathrm{X}}}^\prime \mathrm{Y}^\prime}_{Nk}\rangle
&=\sum_{\mu\in\bar{\mathrm{X}}\mathrm{Y},\nu\in{\bar{\mathrm{X}}}^\prime \mathrm{Y}^\prime}
C^k_{\mu M}\langle\Phi_\mu|\hat{E}_M|\Phi_\nu\rangle C^k_{\nu N} \nonumber\\
&=ELS(\mathrm{YY}^\prime) \cdot HLS(\bar{\mathrm{X}}{\bar{\mathrm{X}}}^\prime) \cdot \textbf{ALSC}^k_{MN},\\
\textbf{ALSC}^k_{MN}&=\sum_{\mu_a\in\bar{\mathrm{X}}\mathrm{Y},\nu_a\in{\bar{\mathrm{X}}}^\prime \mathrm{Y}^\prime}A^k_{\mu_aM} \cdot ALS(\bar{\mathrm{X}}\mathrm{Y},{\bar{\mathrm{X}}}^\prime \mathrm{Y}^\prime)_{\mu_a\nu_a} \cdot A^k_{\nu_aN}  \nonumber\\
&=[(\mathbf{A}^k)^\dag \cdot \textbf{ALS} \cdot \mathbf{A}^k]_{MN},
\end{align}
where $\textbf{ALSC}^k$ represents the assembled matrix of internally contracted partial loops for each state $k$.

\section{Results and Discussion}\label{Results}
The spin-free exact two-component (sf-X2C) relativistic Hamiltonian \cite{X2CSOC1,X2CSOC2}
was employed throughout to account for scalar relativistic effects. The iCISCF(2)\cite{iCISCF} calculations, which
took a single parameter ($C_{\mathrm{min}}$) to prune away unimportant CSFs during the iterations, were initiated and governed by
the iCAS (imposed selection of CAS) approach\cite{iCAS}. Upon convergence,
the contributions of the discarded CSFs (within the CAS) to the energy were accounted for
via ENPT2, leading to $E^{(0)}_k$.
Those CSFs in the iCISCF wavefunctions $\Psi_k^{(0)}$ were further discarded if their coefficients
are smaller in absolute value than $P_{\mathrm{min}}$ for all the target states.
The Dyall Hamiltonian was then diagonalized in the resulting $\tilde{P}$ space, leading to
$\tilde{\Psi}_k^{(0)}$ and $\tilde{E}^{(0)}_k$ for subsequent SDSPT2s calculations of dynamic correlation energies $E_k^{(2)}$, where the TAODs
were screened as discussed before. The SDSPT2s total energies reported here refer to $E^{(0)}_k+E^{(2)}_k$, which is
to be compared with $\tilde{E}^{(0)}_k+E^{(2)}_k$ as advocated in Ref \citenum{Cu2O2-2020}.
The corresponding NEVPT2s results were obtained automatically.
All calculations were performed with the BDF program package \cite{BDF1,BDF2,BDF3,BDFrev2020}
on a computer equipped with 2 Intel(R) Xeon(R) Platinum 8375C CPUs (64 cores in total) and 1 TB memory.

\subsection{Cutoff Thresholds $C_{\mathbf{min}}$, $P_{\mathbf{min}}$, $Q_{\mathbf{min}}$, and DVD}\label{Truncation}
A simplified model of heme (denoted as Fe$^{\mathrm{II}}$L$_2$ with L being N$_2$C$_3$H$_4$)\cite{heme2014,heme2017} was first taken as
a showcase to reveal the effectiveness of the cutoff thresholds, including $C_{\mathbf{min}}$, $P_{\mathbf{min}}$, $Q_{\mathbf{min}}$, and DVD
(see the end of Sec. \ref{CI-subspace}).
The cartesian coordinates were taken from Ref. \citenum{heme2014} (see also Tables S2--S4 in the Supporting Information).
The iCISCF(14,17) calculations were performed for the $^5A_g$, $^3B{_{3g}}$, and $^1A_g$ states separately.
As shown in Fig. S1, the active orbitals include Fe $3d4s4d$ and two $\sigma$, two $\pi$, and two $\pi^\ast$ orbitals
composed of N $2p_{x,y}$ in the molecular plane.
The $1s$ orbital of C and N as well as the $1s2s2p$ orbitals of Fe were frozen in the SDSPT2s calculations.
The def2-TZVP\cite{def2-basis} and corresponding auxiliary basis sets\cite{def2-basis-RI} for the resolution of identity (RI) approximation
were employed in both the iCISCF and SDSPT2s calculations.

It is first seen from Table \ref{iCISCF-cmin} that iCISCF(2) converges steadily to CASSCF as $C_{\mathbf{min}}$ decreases. It
can also be seen from Table \ref{SDSPT2-cmin} that the SDSPT2s energies are only weakly dependent on $C_{\mathbf{min}}$, and
a value of $10^{-4}$ for $C_{\text{min}}$ is already sufficient, which is in line with the previous findings\cite{iCISCF}.
In this case, the number of CSFs in the reference space is merely a few of ten thousands of that of CAS(14,17) as shown in Table \ref{SDSPT2-Nq}.
It is also of interest to see from Table \ref{SDSPT2-cmin} that the DVD approximation does not affect discernibly the spin gaps,
even though ca. 25\% of the TAODs have been pruned away (see Table \ref{SDSPT2-Nq}).
It appears that the smaller the $C_{\mathbf{min}}$, the less effective the DVD. This is not surprising, for a smaller $C_{\mathbf{min}}$
implies more reference oCFGs and hence more survival intermediate nodes before the DVD condition is met.
The results in Table \ref{SDSPT2-pmin} merely show that SDSPT2s is rather insensitive to $P_{\mathrm{min}}$,
and $P_{\mathrm{min}}=10^{-3}$ is already sufficient.
On top of this, a value of $10^{-5}$ for $Q_{\text{min}}$ can be invoked to further screen the TAODs, as can be seen from Table \ref{SDSPT2-qmin}.
Note in passing that the use of $Q_{\text{min}}$ eliminates a large number of individual excited CSFs but does not reduce
the number of ICCs very much. This explains the robustness of such integral-based screening. Unless otherwise stated,
$(C_{\text{min}}=10^{-4}$, $P_{\text{min}}=10^{-3}$, DVD, $Q_{\text{min}}=10^{-5})$ will be taken as the default setting in subsequent calculations.
With this setting, SDSPT2s yields 4.1 and 38.7 kcal/mol for
the gaps between $^5A_g$ and $^3B_{3g}$ and between $^5A_g$ and $^1A_g$
 of Fe$^{\mathrm{II}}$L$_2$, respectively, which  are
 in good agreement with those (3.8 and 40.7 kcal/mol) by CCSD\cite{heme2014}.

\begin{threeparttable}
  \caption{iCISCF(14,17) energies (+1725.0 $E_h$) of Fe$^{\mathrm{II}}$L$_2$ with sf-X2C/def2-TZVP}
  \tabcolsep=18pt
  \scriptsize
  \begin{tabular}[t]{@{}cccccccccccc@{}}
   \hline\hline
  $C_{\text{min}}$  & $E^{(0)}_{var}(^5A_g)$\tnote{a} & $E^{(0)}(^5A_g)$\tnote{b}   & $E^{(0)}_{var}(^3B_{3g})$\tnote{a} & $E^{(0)}(^3B_{3g})$\tnote{b} & $E^{(0)}_{var}(^1A_g)$\tnote{a} & $E^{(0)}(^1A_g)$\tnote{b}  \\
   \hline
$1.0\times10^{-3}$ &	-0.489530 &	-0.494016 &	-0.465730 &	-0.473060 &	-0.418774 &	-0.425355 \\
$5.0\times10^{-4}$ &	-0.491718 &	-0.494107 &	-0.469025 &	-0.473359 &	-0.421739 &	-0.425385 \\
$1.0\times10^{-4}$ &	-0.493753 &	-0.494260 &	-0.472586 &	-0.473590 &	-0.423578 &	-0.424461 \\
$5.0\times10^{-5}$ &	-0.493983 &	-0.494275 &	-0.473023 &	-0.473627 &	-0.423964 &	-0.424486 \\
$0.0$              &	-0.494286 &	-0.494286 &	-0.473641 &	-0.473641 &	-0.424498 &	-0.424498 \\
   \hline \hline
  \end{tabular}\label{iCISCF-cmin}
  \begin{tablenotes}
  \item[] $^\mathrm{a}$ $E^{(0)}_{var}$: variational energy; $^\mathrm{b}$ $E^{(0)}$: $E^{(0)}_{var}$ plus inner space ENPT2 correction.
  \end{tablenotes}
\end{threeparttable}
\vspace{1cm}

\begin{threeparttable}
  \caption{Dependence of SDSPT2s energies (+1727.0 $E_h$) of Fe$^{\mathrm{II}}$L$_2$ on $C_{\text{min}}$ and DVD
  [sf-X2C/def2-TZVP, iCISCF(14,17), $P_{\text{min}}=C_{\text{min}}$, $Q_{\text{min}}=0.0$]}
  \tabcolsep=25pt
  \scriptsize
  \begin{tabular}[t]{@{}ccccccccccccc@{}}
   \hline\hline
   $C_{\text{min}}$  & $E(^5A_g)$    & $E(^3B_{3g})$  & $E(^1A_g)$
   & $\triangle E_1$\tnote{a} & $\triangle E_2$\tnote{b} \\
   \hline
   & &$E=E^{(0)}_k+E^{(2)}_k$, \textbf{w/o DVD} &   & &\\
   $1.0\times10^{-3}$ & -0.505898     & -0.498242    & -0.437672  & 4.8 	&42.8 \\
   $5.0\times10^{-4}$ & -0.506027     & -0.498808 	 & -0.437506  & 4.5 	&43.0 \\
   $1.0\times10^{-4}$ & -0.505800     & -0.499096    & -0.444090  & 4.2 	&38.7 \\
   $5.0\times10^{-5}$ & -0.505916     & -0.499364    & -0.444459  & 4.1 	&38.6 \\
   & & $E=E^{(0)}_k+E^{(2)}_k$, \textbf{w/ DVD} &   & &\\
   $1.0\times10^{-3}$ & -0.505526     & -0.497677    & -0.436842  & 4.9 	&43.1 \\
   $5.0\times10^{-4}$ & -0.505852     & -0.498517    & -0.437150  & 4.6 	&43.1 \\
   $1.0\times10^{-4}$ & -0.505751     & -0.499040    & -0.444233  & 4.2 	&38.6 \\
   $5.0\times10^{-5}$ & -0.505895     & -0.499348    & -0.444503  & 4.1 	&38.5 \\
   & &$E=\tilde{E}^{(0)}_k+E^{(2)}_k$, \textbf{w/o DVD} &   & &\\
   $1.0\times10^{-3}$ & -0.501337      & -0.491265     & -0.431151   & 6.3  	&44.0  \\
   $5.0\times10^{-4}$ & -0.503629      & -0.494369     & -0.434203   & 5.8  	&43.6  \\
   $1.0\times10^{-4}$ & -0.505372      & -0.498232     & -0.443284   & 4.5  	&39.0  \\
   $5.0\times10^{-5}$ & -0.505660      & -0.498849     & -0.443993   & 4.3  	&38.7  \\
   & & $E=\tilde{E}^{(0)}_k+E^{(2)}_k$, \textbf{w/ DVD} &   & &\\
   $1.0\times10^{-3}$ & -0.500965      & -0.490700     & -0.430335   & 6.4  	&44.3  \\
   $5.0\times10^{-4}$ & -0.503454      & -0.494082     & -0.433846   & 5.9  	&43.7  \\
   $1.0\times10^{-4}$ & -0.505291      & -0.498184     & -0.443422   & 4.5  	&38.8  \\
   $5.0\times10^{-5}$ & -0.505639      & -0.498832     & -0.444037   & 4.3  	&38.7  \\
   \hline \hline
  \end{tabular}\label{SDSPT2-cmin}
  \begin{tablenotes}
  \item[] $^\mathrm{a}$ energy gap (in kcal/mol) between $^5A_g$ and $^3B_{3g}$; $^\mathrm{b}$ energy gap (in kcal/mol) between $^5A_g$ and $^1A_g$.
  \end{tablenotes}
\end{threeparttable}
\vspace{1cm}

\begin{threeparttable}
  \caption{Numbers of parameters (as function of $C_{\text{min}}$ and DVD) in SDSPT2s calculations of Fe$^{\mathrm{II}}$L$_2$
  [sf-X2C/def2-TZVP, iCISCF(14,17), $P_{\text{min}}=C_{\text{min}}$, $Q_{\text{min}}=0.0$]}.
  \tabcolsep=21pt
  \scriptsize
  \begin{tabular}[t]{@{}ccccccccccccc@{}}
   \hline\hline
   $C_{\text{min}}$ &  $N_R/N_{CAS}\times 10^4$ \tnote{a} & $N_1$ \tnote{b} & $N_2/N_1$ (\%)\tnote{c} & $N_3$ \tnote{d} & $N_4/N_3$ (\%)\tnote{e} \\
   \hline
         & & & $^5A_g$ & &\\
   $1.0\times10^{-3}$ & 0.2 &   1473684 & 69.8 &  29652784 & 73.1 \\
   $5.0\times10^{-4}$ & 0.4 &   4643615 & 70.9 &  87211642 & 75.1 \\
   $1.0\times10^{-4}$ & 1.44 &  13096536 & 71.6 & 251826337 & 74.7 \\
   $5.0\times10^{-5}$ & 2.33 &  20281378 & 73.7 & 383337864 & 78.1 \\
         & & & $^3B_{3g}$ & &\\
   $1.0\times10^{-3}$ & 0.2 &   2555799 & 69.5 &  40553046 & 72.0 \\
   $5.0\times10^{-4}$ & 0.4 &   8783390 & 72.6 & 121732995 & 73.6 \\
   $1.0\times10^{-4}$ & 1.6 &  23520762 & 76.6 & 328815443 & 77.4 \\
   $5.0\times10^{-5}$ & 2.9 &  39305048 & 77.7 & 539232278 & 79.9 \\
         & & & $^1A_g$ & &\\
   $1.0\times10^{-3}$ & 0.6 &   2630783 & 66.4 &  20620925 & 65.8 \\
   $5.0\times10^{-4}$ & 1.1 &   6460877 & 66.8 &  48328424 & 64.5 \\
   $1.0\times10^{-4}$ & 4.5 &  21965386 & 74.4 & 168755703 & 75.2 \\
   $5.0\times10^{-5}$ & 7.8 &  38131568 & 75.5 & 286693520 & 77.0 \\
   \hline \hline
  \end{tabular}\label{SDSPT2-Nq}
  \begin{tablenotes}
  \item[]
  $^\mathrm{a}$ $N_R$: number of selected reference CSFs; $N_{CAS}$: number of CAS reference CSFs
  (13009252, 18804520, and 9690780 for $^5A_g$, $^3B_{3g}$, and $^1A_g$, respectively);
  $^\mathrm{b}$ $N_1$: number of $\tilde{Q}_1(\bar{\mathrm{D}}\mathrm{V})$ without DVD;
  $^\mathrm{c}$ $N_2$: number of $\tilde{Q}_1(\bar{\mathrm{D}}\mathrm{V})$ with DVD;
  $^\mathrm{d}$ $N_3$: number of $\tilde{Q}_1(\bar{\mathrm{V}}\mathrm{D})$ without DVD;
  $^\mathrm{e}$ $N_4$: number of $\tilde{Q}_1(\bar{\mathrm{V}}\mathrm{D})$ with DVD.
  \end{tablenotes}
\end{threeparttable}
\vspace{1cm}

\begin{threeparttable}
  \caption{Dependence of SDSPT2s energies (+1727.0 $E_h$) of Fe$^{\mathrm{II}}$L$_2$ on $P_{\mathrm{min}}$
  [sf-X2C/def2-TZVP, iCISCF(14,17), $C_{\text{min}}=10^{-4}$, $Q_{\text{min}}=0.0$, DVD]}
  \tabcolsep=27pt
  \scriptsize
  \begin{tabular}[t]{@{}ccccccccccccc@{}}
   \hline\hline
  $P_{\text{min}}$  & $E(^5A_g)$    & $E(^3B_{3g})$  & $E(^1A_g)$
   & $\triangle E_1$\tnote{a} & $\triangle E_2$\tnote{b} \\
   \hline
   & & & $E=E^{(0)}_k+E^{(2)}_k$ & & \\
   $1.0\times10^{-3}$ & -0.506706     & -0.500335    & -0.445045  & 4.0 	&38.7 \\
   $5.0\times10^{-4}$ & -0.505963     & -0.499160  	 & -0.444192  & 4.3 	&38.8 \\
   $1.0\times10^{-4}$ & -0.505751     & -0.499040    & -0.444233  & 4.2 	&38.6 \\
   & & & $E=\tilde{E}^{(0)}_k+E^{(2)}_k$ & & \\
   $1.0\times10^{-3}$ & -0.503998 	& -0.495798 &	-0.440368 &	5.1 &	39.9 \\
   $5.0\times10^{-4}$ & -0.504875 	& -0.497441 &	-0.442269 &	4.7 &	39.3 \\
   $1.0\times10^{-4}$ & -0.505290 	& -0.498184 &	-0.443423 &	4.5 &	38.8 \\
   \hline \hline
  \end{tabular}\label{SDSPT2-pmin}
  \begin{tablenotes}
  \item[] $^\mathrm{a}$ energy gap (in kcal/mol) between $^5A_g$ and $^3B_{3g}$; $^\mathrm{b}$ energy gap (in kcal/mol) between $^5A_g$ and $^1A_g$.
  \end{tablenotes}
\end{threeparttable}
\vspace{1cm}

\begin{threeparttable}
  \caption{Numbers of parameters and energies (as function of $Q_{\mathrm{min}}$) in SDSPT2s calculations of Fe$^{\mathrm{II}}$L$_2$
  [sf-X2C/def2-TZVP, iCISCF(14,17), $C_{\text{min}}=10^{-4}$, $P_{\text{min}}=10^{-3}$, DVD]}
  \tabcolsep=18pt
  \scriptsize
  \begin{tabular}[t]{@{}lcccccccccccc@{}}
   \hline\hline
   &  \multicolumn{4}{c}{$Q_{\text{min}}$} \\
   \cline{2-5}
   &  $1.0\times10^{-3}$ & $1.0\times10^{-4}$ &  $1.0\times10^{-5}$  & 0.0 \\
   \hline
   $N_{\bar{\mathrm{D}}\mathrm{V}}$/$\bar{N}_{(\bar{\mathrm{D}}\mathrm{V})}$ of $^5A_g$\tnote{a}    & 162569/4049    & 348874/12443  & 1139368/17236   & 3835035/18391  \\
   $N_{\bar{\mathrm{V}}\mathrm{D}}$/$\bar{N}_{(\bar{\mathrm{V}}\mathrm{D})}$ of $^5A_g$\tnote{b}    & 3348671/35627  &3964792/105844 & 10823677/199235 &78491786/236659 \\
   $N_{\bar{\mathrm{D}}\mathrm{V}}$/$\bar{N}_{(\bar{\mathrm{D}}\mathrm{V})}$ of $^3B_{3g}$\tnote{a} & 162183/4829    & 473459/13820  & 2081959/17716   & 6143488/18427  \\
   $N_{\bar{\mathrm{V}}\mathrm{D}}$/$\bar{N}_{(\bar{\mathrm{V}}\mathrm{D})}$ of $^3B_{3g}$\tnote{b} & 3895457/35512  &4788344/114775 & 13692840/202556 &92404719/237653 \\
   $N_{\bar{\mathrm{D}}\mathrm{V}}$/$\bar{N}_{(\bar{\mathrm{D}}\mathrm{V})}$ of $^1A_g$\tnote{a}    & 220640/5644    & 560727/14516  &
   2215580/17826   & 4770647/18384  \\
   $N_{\bar{\mathrm{V}}\mathrm{D}}$/$\bar{N}_{(\bar{\mathrm{V}}\mathrm{D})}$ of $^1A_g$\tnote{b}    & 1548920/36238  &2330557/117176 &
   8106330/202719  &34877101/234687\\
   & &  $E=E^{(0)}_k+E^{(2)}_k$ & & \\
   $E(^5A_g)$\tnote{c}     & -0.502386 &	-0.505569 &	-0.506566 &	-0.506706 \\
   $E(^3B_{3g})$\tnote{c}  & -0.492765 &	-0.498354 &	-0.500099 &	-0.500335 \\
   $E(^1A_g)$\tnote{c}     & -0.438914 &	-0.443763 &	-0.444933 &	-0.445045 \\
   $\triangle E_1$\tnote{d} & 6.0  &	4.5 &	4.1 &	4.0  \\
   $\triangle E_2$\tnote{e} & 39.8 &	38.8&	38.7&	38.7 \\
   & & $E=\tilde{E}^{(0)}_k+E^{(2)}_k$ & & \\
   $E(^5A_g)$\tnote{c}      & -0.499678 & -0.502862 &	-0.503859 &	-0.503998 \\
   $E(^3B_{3g})$\tnote{c}   & -0.488228 & -0.493817 &	-0.495562 &	-0.495798 \\
   $E(^1A_g)$\tnote{c}      & -0.434237 & -0.439087 &	-0.440256 &	-0.440368 \\
   $\triangle E_1$\tnote{d} & 7.2 &   5.7  &5.2 	&5.1 \\
   $\triangle E_2$\tnote{e} & 41.1&   40.0 &39.9 	&39.9\\
   \hline \hline
  \end{tabular}\label{SDSPT2-qmin}
  \begin{tablenotes}
  \item[]
  $^\mathrm{a}$ $N_{\bar{\mathrm{D}}\mathrm{V}}$: number of $\tilde{Q}_1(\bar{\mathrm{D}}\mathrm{V})$; $\bar{N}_{(\bar{\mathrm{D}}\mathrm{V})}$: number of $\bar{\Phi}^{\bar{\mathrm{D}}\mathrm{V}}_{Mk}$ (see Eq. \eqref{ic-definition});
  $^\mathrm{b}$ $N_{\bar{\mathrm{V}}\mathrm{D}}$: number of $\tilde{Q}_1(\bar{\mathrm{V}}\mathrm{D})$; $\bar{N}_{(\bar{\mathrm{V}}\mathrm{D})}$:  number of $\bar{\Phi}^{\bar{\mathrm{V}}\mathrm{D}}_{Mk}$ (see Eq. \eqref{ic-definition});
  $^\mathrm{c}$ total energy shifted by 1727.0 $E_h$;
  $^\mathrm{d}$ energy gap (in kcal/mol) between $^5A_g$ and $^3B_{3g}$; $^\mathrm{e}$ energy gap (in kcal/mol) between $^5A_g$ and $^1A_g$.
  \end{tablenotes}
\end{threeparttable}

\subsection{Computational Resources}
To see the gain in efficiency of SDSPT2s over SDSPT2, we carried out calculations for the ground and lowest
triplet states of anthracene with the cc-pVDZ basis set\cite{cc-pVZ-basis} and  $D_{2\textrm{h}}$ symmetry.
The geometry of anthracene was taken from Ref. \citenum{ASCISCF2}.
The (14e, 14o) active space is composed of the valence $\pi$ and $\pi^\ast$ orbitals. As can be seen from
Table \ref{time-mem}, except for the construction of the sub-DRTs, all other steps
became significantly cheaper by incorporating the cutoffs, leading to an overall gain in efficiency by more than one order of magnitude,
yet without sacrificing the accuracy. In particular, the memory requirement is also reduced substantially.

\begin{threeparttable}
  \caption{Detailed information for CAS(14,14)-SDSPT2/cc-pVDZ calculations of the $S_0$ and $T_1$ states of anthracene}
  \tabcolsep=23pt
  \scriptsize
  \begin{tabular}[t]{@{}lccccccccccc@{}}
   \hline\hline
   &  $S_0$ \tnote{a} &  $S_0$ \tnote{b} & $T_1$ \tnote{a} & $T_1$ \tnote{b} \\
   \hline
   $T_{DRT}/s$ \tnote{c}                                            & 147  &      15 & 	231 	& 19     \\
   $T_{\Psi^{(0)}_k}/s$ \tnote{d}                                   & 3    &     615 & 	14   	& 2693   \\
   $T_{\bar{\Phi}^{\bar{\mathrm{X}}\mathrm{Y}}_{Mk}}/s$ \tnote{e}   & 178  &    1129 & 	632 	& 3422   \\
   $T_{\Psi^{(0)}_{q,k}}/s$ \tnote{f}                               & 1390 & 	17471 & 	4063 	& 36481  \\
   $T_{\Xi^{(1)}_k}/s$ \tnote{g}                                    & 123  &    4482 & 	430 	& 10646  \\
   $T_{\Theta^{(2)}_k}/s$ \tnote{h}                                 & 124  &    5609 & 	438 	& 14051  \\
   $T_{\tilde{\textbf{S}}}/s$ \tnote{i}                             & 24   &     321 & 	58 	    & 636    \\
   $T_{\tilde{\textbf{H}}}/s$ \tnote{j}                             & 244  &    9030 & 	863 	& 21339  \\
   $T_{tot}/s$ \tnote{k}                                            & 2234 & 	38671 & 	6729 	& 89287  \\
   $M_{max}/\mathrm{GB}$ \tnote{l}  &    0.5    &    5.8        &     0.9      &    11.4 \\
   $N_R$ \tnote{m}      & 6725  &       691160 &  17158  &       1252895 \\
   $N_q$ \tnote{n}      & 4.751$\times 10^{10}$  &  6.134$\times 10^{12}$ & 2.143$\times 10^{11}$ &   1.327$\times 10^{13}$ \\
   $E$ \tnote{o}  & -0.834178 (-0.829521) &  -0.829089  & -0.763068 (-0.756254) &    -0.755786 \\
   $\triangle E$ \tnote{p} &     0.0 (0.0)      &     0.0       &  44.6 (46.0) &  46.0  \\
   \hline \hline
  \end{tabular}\label{time-mem}
  \begin{tablenotes}
  \item[]
  $^\mathrm{a}$ SDSPT2s ($C_{\text{min}}=10^{-4}$, $P_{\text{min}}=10^{-3}$, $Q_{\text{min}}=10^{-5}$, DVD);
  $^\mathrm{b}$ SDSPT2;
  $^\mathrm{c}$ time for constructing (reduced) sub-DRTs (see \textbf{Algorithm 1} or \textbf{Algorithm S1});
  $^\mathrm{d}$ time for calculating $\Psi^{(0)}_k$;
  $^\mathrm{e}$ time for calculating ICC $\bar{\Phi}^{\bar{\mathrm{X}}\mathrm{Y}}_{Mk}$ [see Eq. (\ref{ic-definition});
  including generation of $\tilde{Q}_1$ [see Eq. (\ref{Qmin})] in the case of SDSPT2s];
  $^\mathrm{f}$ time for calculating $\Psi^{(0)}_{q,k}$ [see Eq.(\ref{DyallDiag})];
  $^\mathrm{g}$ time for calculating $\Xi^{(1)}_k$ [see Eq. (\ref{PTWF1def})];
  $^\mathrm{h}$ time for calculating $\Theta^{(2)}_k$ [see Eq. (\ref{LanczosVec})];
  $^\mathrm{i}$ time for constructing $\tilde{\textbf{S}}$ [see Eq. (\ref{Smat})];
  $^\mathrm{j}$ time for constructing $\tilde{\textbf{H}}$ [see Eq. (\ref{Hmat2})];
  $^\mathrm{k}$ total time;
  $^\mathrm{l}$ peak memory;
  $^\mathrm{m}$ dimension of $\tilde{P}$ or $P$;
  $^\mathrm{n}$ dimension of $\tilde{Q}_1$ or $Q$;
  $^\mathrm{o}$ total energy shifted by 537.0 $E_h$; in parentheses are $\tilde{E}^{(0)}_k+E^{(2)}_k$;
  $^\mathrm{p}$ relative energy (in kcal/mol) with respective to $S_0$; in parentheses are derived from $\tilde{E}^{(0)}_k+E^{(2)}_k$.
  \end{tablenotes}
\end{threeparttable}

\subsection{Size Consistency}\label{Cr2}
It has been shown\cite{SDSrev} that
the size-consistency errors of SDSPT2 can be largely removed by using the Pople correction\cite{Size4}.
To see if this is also the case for SDSPT2s, we calculated the three lowest
singlet and three lowest triplet states of the anthracene...Rg complexes, with Rg (= He, Ne, Ar, Kr) set to 100 ~\AA~ away from the center of anthracene.
Three-state-averaged (SA3) CAS/iCISCF(14,14) calculations were first performed with the cc-pVDZ basis set\cite{cc-pVZ-basis} and  $C_{2\textrm{v}}$ symmetry.
It can be seen from Table \ref{SizeCons} that the size-consistency errors of SDSPT2s are indeed very much the same as those of CAS-based SDSPT2.
Note in passing that, at variance with the state-specific NEVPT2 (which is strictly size consistent\cite{NEVPT2a}), MS-NEVPT2s/NEVPT2 is also not size consistent but the errors are much smaller than those of SDSPT2s/SDSPT2. Yet, the advantage of SDSPT2s/SDSPT2 lies in the unified treatment of single and multiple states,
which is particularly warranted for situations with near degeneracies\cite{SDSPT2}.

\begin{threeparttable}
  \caption{Size-consistency errors (in meV) of SA3-iCISCF(14,14), MS-NEVPT2s, and SDSPT2s for the excitation energies (relative to anthracene) of the
  low-lying singlet and triplet states of anthracene...Rg (Rg = He, Ne, Ar, Kr set to 100 {\AA} away from the center of anthracene;
  cc-pVDZ; $C_{\text{min}}=10^{-4}$, $P_{\text{min}}=10^{-3}$, $Q_{\text{min}}=10^{-5}$, DVD)}
  \tabcolsep=10pt
  \scriptsize
  \begin{tabular}[t]{@{}ccccccccccccc@{}}
   \hline\hline
   Species & iCISCF & MS-NEVPT2s\tnote{a} & MS-NEVPT2 & SDSPT2s\tnote{a,b} & SDSPT2\tnote{b} \\
   \hline
   & & & Singlet & &\\
   anthracene$\cdots$He $1^1A_1$ & 0.0 &	 0.1 (0.0) &	 0.0  &	 1.9 (2.6)    &	 1.8 \\
   anthracene$\cdots$He $2^1A_1$ & 0.0 &	 2.5 (2.2) &	 2.6  &	 4.0 (3.4)    &	 4.0 \\
   anthracene$\cdots$He $3^1A_1$ & 0.0 &	 2.4 (2.1) &	 1.9  &	 4.4 (7.1)    &	 3.8 \\
   anthracene$\cdots$Ne $1^1A_1$ & 0.0 &	 0.3 (0.0) &	 0.0  & -2.0 (-39.3)  &  -2.3 \\
   anthracene$\cdots$Ne $2^1A_1$ & 0.0 &	 1.9 (1.7) &	 2.6  & -1.7 (-2.3)   &  -1.5 \\
   anthracene$\cdots$Ne $3^1A_1$ & 0.0 &	 2.7 (2.3) &	 1.9  &  1.8 (2.8)    &   1.0 \\
   anthracene$\cdots$Ar $1^1A_1$ & 0.0 &	 0.2 (0.0) &	 0.0  &	30.2 (28.3)   &	29.7 \\
   anthracene$\cdots$Ar $2^1A_1$ & 0.0 &	 2.4 (2.2) &	 2.7  &	31.4 (31.1)   &	30.7 \\
   anthracene$\cdots$Ar $3^1A_1$ & 0.0 &	 2.3 (1.9) &	 2.7  &	33.8 (33.4)   &	33.5 \\
   anthracene$\cdots$Kr $1^1A_1$ & 0.0 &	 0.3 (0.1) &	 0.0  &	95.6 (68.8)   &	94.6 \\
   anthracene$\cdots$Kr $2^1A_1$ & 0.0 &	 2.9 (2.3) &	 2.5  &	97.2 (98.0)   &	94.9 \\
   anthracene$\cdots$Kr $3^1A_1$ & 0.0 &	 2.7 (2.0) &	 1.7  &100.3 (99.4)   &  97.2 \\
   & & & Triplet & & \\
   anthracene$\cdots$He $1^3A_1$ & 0.2 &	-5.3 (0.0) &	 1.3  &	-3.1 (2.2)  &   3.0 \\
   anthracene$\cdots$He $2^3A_1$ & 0.1 &	 0.0 (0.3) &	 2.1  &	 1.7 (2.6)  &   3.8 \\
   anthracene$\cdots$He $3^3A_1$ & 0.0 &	 6.2 (2.4) &	 2.0  &	 7.7 (2.7)  &   3.6 \\
   anthracene$\cdots$Ne $1^3A_1$ & 0.0 &	 1.3 (1.7) &	 1.3  & -1.1 (-1.3) &  -1.4 \\
   anthracene$\cdots$Ne $2^3A_1$ & 0.0 &	 6.2 (3.0) &	 2.1  &  3.8 (0.8)  &  -0.3 \\
   anthracene$\cdots$Ne $3^3A_1$ & 0.0 &	 8.0 (4.1) &	 2.1  &  5.5 (0.5)  &  -0.4 \\
   anthracene$\cdots$Ar $1^3A_1$ & 0.2 &	-6.4 (0.1) &	 1.2  &	24.3 (30.7) &  30.6 \\
   anthracene$\cdots$Ar $2^3A_1$ & 0.1 &	-0.1 (0.5) &	 2.1  &	30.3 (30.7) &  31.7 \\
   anthracene$\cdots$Ar $3^3A_1$ & 0.0 &	 8.8 (3.8) &	 2.1  &	38.0 (32.9) &  30.8 \\
   anthracene$\cdots$Kr $1^3A_1$ & 0.3 &   -12.6 (-1.7) &     1.3  &84.4 (94.9) &  95.5 \\
   anthracene$\cdots$Kr $2^3A_1$ & 0.1 &	-5.6 (-1.4) &	 2.1  &	90.7 (94.9) &  96.4 \\
   anthracene$\cdots$Kr $3^3A_1$ & 0.0 &	 7.6 (2.4) &	 2.2  &101.5 (95.4) &  94.5 \\
   \hline \hline
  \end{tabular}\label{SizeCons}
  \begin{tablenotes}
  \item[a] Results in parentheses were derived from $\tilde{E}^{(0)}_k+E^{(2)}_k$
  \item[b] With Pople correction.
  \end{tablenotes}
 \end{threeparttable}

2
\subsection{PEC of Cr$_2$}\label{Cr2}
\begin{figure}
\centering
\includegraphics[width=1.0\textwidth]{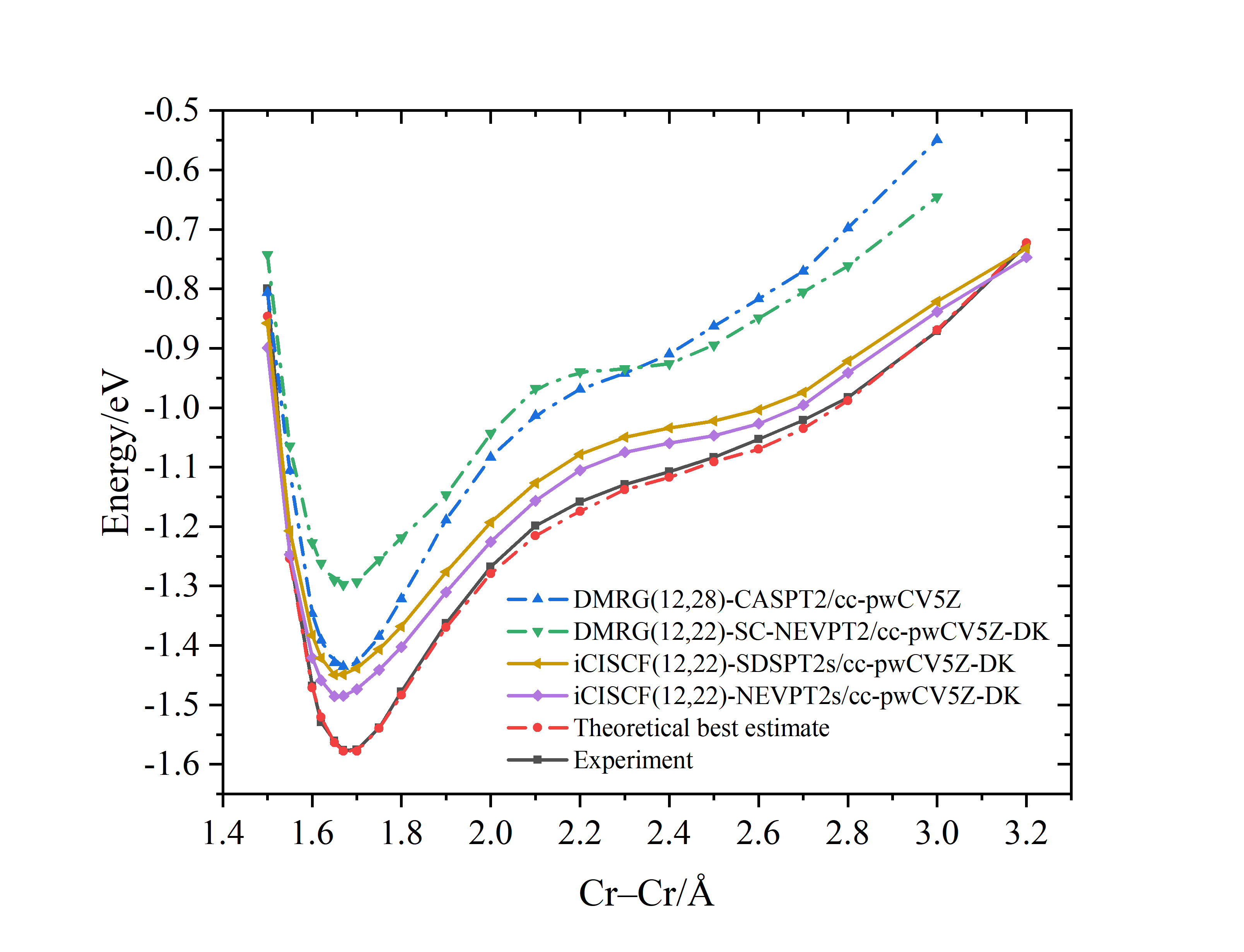}
\caption{Potential energy curves of Cr$_2$ computed by iCISCF(12,22)-SDSPT2s/NEVPT2s [$C_{\text{min}}=10^{-4}$, $P_{\text{min}}=10^{-3}$,
$Q_{\text{min}}=10^{-5}$, DVD],
to compare with those by DMRG(12,22)-SC-NEVPT2\cite{Cr2-2016}, DMRG(12,28)-CASPT2\cite{Cr2-2011},
theoretical best estimate (TBE)\cite{Cr2-2022}, and revised experiment\cite{Cr2-2022}. }
\label{Cr2-PEC}
\end{figure}

The potential energy curve (PEC) of Cr$_2$ has been investigated extensively in the last decades \cite{Cr2-1983,Cr2-1993,Cr2-1994,Cr2-1998,Cr2-2016,Cr2-2011,Cr2-2018,Cr2-2020,Cr2-2022}.
The CAS(12,12) PEC turns out to be repulsive, which is qualitatively wrong. Therefore, it is mandatory to use larger
active spaces, e.g., CAS(12,22)\cite{Cr2-2016} ($3d4s4d$), CAS(12,28)\cite{Cr2-2011} ($3d4s4d4p$), and even CAS(12,42)\cite{Cr2-2018} ($3d4s4d4p4f$)
in multi-reference calculations.
Here, we used the CAS(12,22) active space, in conjunction with the cc-pwCV5Z-DK basis set\cite{cc-pwCV5Z-DK} and $D_{2\textrm{h}}$ symmetry.
The SDSPT2s and NEVPT2s PECs are plotted in Fig. \ref{Cr2-PEC}, along with those by DMRG(12,22)-SC-NEVPT2 (DMRG: density matrix renormalization group;
SC: strong contraction)\cite{Cr2-2016}, DMRG(12,28)-CASPT2\cite{Cr2-2011},
theoretical best estimate (TBE)\cite{Cr2-2022}, and revised experiment\cite{Cr2-2022}. As can be seen
from Fig. \ref{Cr2-PEC}, the present SDSPT2s and NEVPT2s PECs are very close to each other and are notably
better than those by DMRG(12,22)-SC-NEVPT2\cite{Cr2-2016} and DMRG(12,28)-CASPT2\cite{Cr2-2011}.
In particular, our SDSPT2s and NEVPT2s PECs are consistent with the latest TBE and revised experimental results\cite{Cr2-2022}
in the region between 2.0 and 2.4 {\AA}, where the DMRG(12,22)-SC-NEVPT2 PEC exhibits a plateau\cite{Cr2-2016}.
The good quality of our SDSPT2s and NEVPT2s PECs is supported by the proximity and parallelity of the variational energies of iCISCF(12,22)
and reduced space $\tilde{P}$, with the nonparallelity of their differences
being only 0.22  mE$_h$  for the whole interatomic distances
(see Table S5 in the Supporting Information). Moreover, the weights of
the CSFs of $\tilde{P}$ are all more than 99\% of those of iCISCF(12,22). As an additional indicator,
the size-consistent errors of both SDSPT2s and NEVPT2s are less than 0.5 kcal/mol.

The spectroscopic constants of Cr$_2$, which were obtained by fitting the PEC between 1.55 and 1.8 {\AA}, are given in Table \ref{Cr2-spec}.
They are in fairly good agreement with
the previous\cite{Cr2-2011,Cr2-2016,Cr2-2018} and experimental\cite{Cr2-1983,Cr2-1993,Cr2-1998} results.


\begin{threeparttable}
  \caption{Spectroscopic constants of Cr$_2$ by various MR methods}
  \tabcolsep=25pt
  \scriptsize
  \begin{tabular}[t]{@{}lcccccccccccc@{}}
   \hline\hline
      & $R_e$ ({\AA}) & $D_e$ (eV) & $\omega _e$ (cm$^{-1}$) \\
   \hline
   DMRG(12,28)-CASPT2/cc-pwCV5Z\tnote{a} & 1.681 & 1.610          & 470 \\
   DMRG(12,22)-SC-NEVPT2/CBS\tnote{b} & 1.655 & 1.435          & 469 \\
   iCISCF(12,22)-NEVPT2s/cc-pwCV5Z-DK\tnote{c} & 1.658 & 1.486          & 469  \\
   iCISCF(12,22)-SDSPT2s/cc-pwCV5Z-DK\tnote{c} & 1.659 & 1.449          & 462  \\
   DMRG(12,42)-ec-MRCISD+Q/ANO-RCC-VQZP\tnote{d} & 1.71 & 1.62          & 479 \\
   Theoretical best estimate\tnote{e} & 1.685 & 1.58$\pm$0.02 & 495 \\
   Experiment             & 1.679\tnote{f} & 1.47(5)\tnote{g}, 1.56$\pm$0.06\tnote{h}   & 481\tnote{g}  \\
   \hline \hline
  \end{tabular}\label{Cr2-spec}
  \begin{tablenotes}
  \item[]
  $^\mathrm{a}$ Ref. \citenum{Cr2-2011};
  $^\mathrm{b}$ Ref. \citenum{Cr2-2016};
  $^\mathrm{c}$ This work;
  $^\mathrm{d}$ Ref. \citenum{Cr2-2018};
  $^\mathrm{e}$ Ref. \citenum{Cr2-2022};
  $^\mathrm{f}$ Ref. \citenum{Cr2-1983};
  $^\mathrm{g}$ Ref. \citenum{Cr2-1993};
  $^\mathrm{h}$ Ref. \citenum{Cr2-1998}.
  \end{tablenotes}
\end{threeparttable}

\subsection{[Cu$_2$O$_2$]$^{2+}$ core}\label{Cu2O2}
\begin{figure}
\centering
\includegraphics[width=1.0\textwidth]{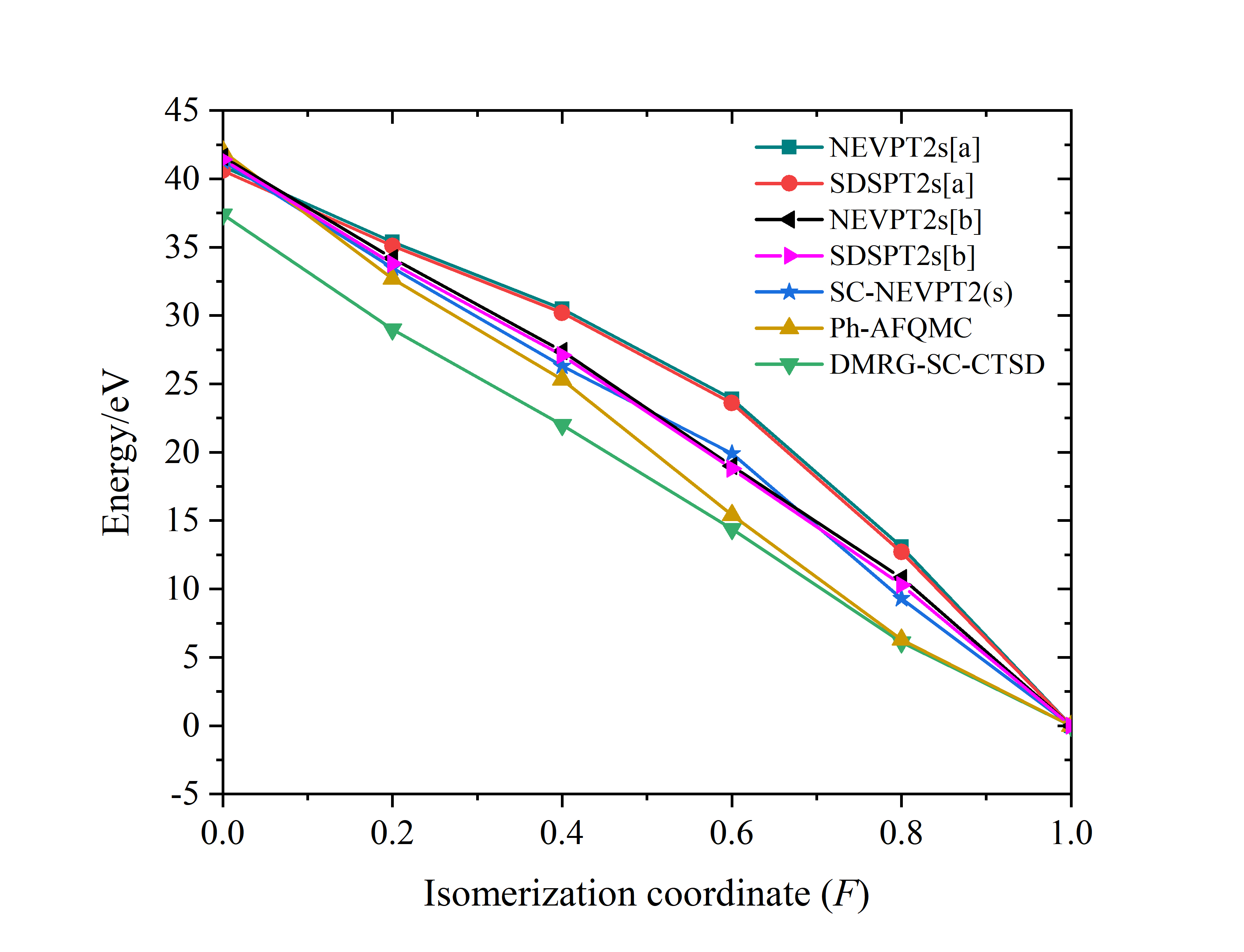}
\caption{Relative energies of [Cu$_2$O$_2$]$^{2+}$ isomers by iCISCF(28,32)-SDSPT2s/NEVPT2s, to compare with those by
HCISCF(28,32)-SC-NEVPT2(s)\cite{Cu2O2-2020},
DMRG(28,32)-SC-CTSD\cite{Cu2O2-2010}, and
Ph-AFQMC (52e results)\cite{Cu2O2-2019}. SDSPT2s[a]/NEVPT2s[a] and SDSPT2s[b]/NEVPT2s[b] were derived from $E^{(0)}_k+E^{(2)}_k$ and $\tilde{E}^{(0)}_k+E^{(2)}_k$, respectively. }
\label{Cu2O2-fig}
\end{figure}

As a bi-copper system to activate O$_2$, [Cu$_2$O$_2$]$^{2+}$ has been a popular model for validating quantum chemical methods.
The isomerization energy between the bis($\mu$-oxo) (\emph{F} = 0.0) and $\mu$-$\eta^2:\eta^2$-peroxo (\emph{F} = 1.0) isomers has been studied extensively by various approaches\cite{Cu2O2-2006,Cu2O2-2008,Cu2O2-2010,Cu2O2-2016,Cu2O2-2020,RR-MRCI,Cu2O2-2019}.
To compare with the previous results by DMRG-SC-CTSD\cite{Cu2O2-2010} and stochastic SC-NEVPT2 (SC-NEVPT2(s))\cite{Cu2O2-2020},
the same active space CAS(28,32) consisting of Cu $3d4d$ and O $2p3p$ was employed here, in conjunction with the ANO-RCC-VQZP basis sets\cite{ANO-RCC-MG,ANO-RCC-TM} and  $D_{2\textrm{h}}$ symmetry. The cartesian coordinates of different [Cu$_2$O$_2$]$^{2+}$ structures were taken from Ref. \citenum{Cu2O2-2006} (see also Table S6 in the Supporting Information). The relative energies calculated by iCISCF(28,32)-NEVPT2s/SDSPT2s are presented in Table \ref{Cu2O2-tab}.
The iCISCF(28,32) results show a shallow well at \emph{F} = 0.8, indicating that accounting only for static correlation is insufficient.
The NEVPT2s and SDSPT2s calculations yield nearly identical results, which are higher by 4 kcal/mol than those by SC-NEVPT2(s)\cite{Cu2O2-2020}
for the structures between  \emph{F} = 0.4 to \emph{F}=0.8. It turns out that such discrepancy stems from
the different assemblies of the SDSPT2s/NEVPT2s and SC-NEVPT2(s) energies,
 $E^{(0)}_k+E^{(2)}_k$ vs. $\tilde{E}^{(0)}_k+E^{(2)}_k$. If the latter\cite{Cu2O2-2020} is also used in SDSPT2s/NEVPT2s,
 the discrepancy of 4 kcal/mol disappears (see values in brackets in Table \ref{Cu2O2-tab}), and the results
 become even closer to those by DMRG-SC-CTSD\cite{Cu2O2-2010} and especially to those by Ph-AFQMC (phaseless auxiliary field
quantum Monte Carlo)\cite{Cu2O2-2019}, see Fig. \ref{Cu2O2-fig}. However, the error compensations between $\tilde{E}^{(0)}_k$ (positive error)
and $E^{(2)}_k$ (negative error)
does not hold in general. For instance, with this combination, the SDSPT2s dissociation energy of Cr$_2$ is only
1.11 eV, which is much smaller than that (1.45 eV) reported in Table \ref{Cr2-spec}.


\begin{threeparttable}
  \caption{Energies (in kcal/mol) of [Cu$_2$O$_2$]$^{2+}$ relative to $\mu$-$\eta^2:\eta^2$-peroxo (\emph{F} = 1.0) }
  \tabcolsep=16pt
  \scriptsize
  \begin{tabular}[t]{@{}lcccccccccccc@{}}
   \hline\hline
   Method  & \emph{F}=0.0 & \emph{F}=0.2 & \emph{F}=0.4 & \emph{F}=0.6 & \emph{F}=0.8 & \emph{F}=1.0 \\
   \hline
iCISCF\tnote{a,b}                   &  18.5    &  11.1   &  5.6    &  1.1     & -1.3   & 0.0 \\
NEVPT2s\tnote{a,b,c}                   &  40.9 [41.6]    &  35.4 [34.2]   & 30.5 [27.4]    & 23.9 [19.0]     & 13.1 [10.8]   & 0.0 [0.0] \\
SDSPT2s\tnote{a,b,c}                   &  40.6 [41.4]    &  35.1 [33.8]   & 30.2 [27.1]    & 23.6 [18.8]     & 12.7 [10.3]   & 0.0 [0.0] \\
SC-NEVPT2(s)\tnote{a,d}    &  41.3(8) &  33.5(9)&  26.3(9)&  19.9(8) & 9.3(9) & 0.0 \\
DMRG-SC-CTSD\tnote{a,e}    &  37.4    &  29.0   &  22.0   &  14.4    &  6.1   & 0.0 \\
Ph-AFQMC\tnote{f}                   &  42.0 & 32.7 & 25.3 &15.4 &6.3 &0.0 \\
  \hline \hline
  \end{tabular}\label{Cu2O2-tab}
  \begin{tablenotes}
  \item[] $^\mathrm{a}$ Active space (28e, 32o); $^\mathrm{b}$ $C_{\text{min}}=10^{-4}$; $^\mathrm{c}$ $P_{\text{min}}=10^{-3}$, $Q_{\text{min}}=10^{-5}$, DVD; values in brackets were derived from $\tilde{E}^{(0)}_k+E^{(2)}_k$;
  $^\mathrm{d}$ Ref.\citenum{Cu2O2-2020}; $^\mathrm{e}$ Ref.\citenum{Cu2O2-2010}; $^\mathrm{f}$ adapted from Ref.\citenum{Cu2O2-2019} (52e results).
  \end{tablenotes}
\end{threeparttable}

\subsection{Cobalt Tropocoronand Complex}\label{CoTCNO}
Transition metal (TM) tropocoronand (TC) nitrosyl complexes, [TM(TC-n,n)(NO)],
have unusual geometric and electronic structures as function of the number of the
alkylidene linkers. As one of the simplest representatives, the cobalt nitrosyl complex [Co(TC-3,3)(NO)] (see Fig. S2 in the Supporting Information)
has been investigated both experimentally\cite{CoTC-1998,CoTCNO2} and theoretically\cite{CoTCNO2,CoTCNO3}.
The early assignment\cite{CoTC-1998} of its ground state to be paramagnetic (which is different from its
homologues with larger $n$) was disapproved by a later density functional theory (DFT) and experimental study\cite{CoTCNO2},
which shows that the ground state of [Co(TC-3,3)(NO)] is diamagnetic.
Interestingly, two distinct triplet states, denoted as $T_1$ and $T_2$, have been obtained by DFT optimizations\cite{CoTCNO2}
(see Tables S7--S9  in the Supporting Information for their cartesian coordinates).
While the DFT reassignment of the ground state of [Co(TC-3,3)(NO)] was confirmed by subsequent DMRG(22,22)-SC-NEVPT2 calculations\cite{CoTCNO3},
the underlying spin coupling mechanisms for the triplet states were interpreted differently.
Therefore,  we decided to carry out iCISCF(22,22)-SDSPT2s calculations with
the def2-TZVP\cite{def2-basis} and corresponding auxiliary basis sets for the RI approximation\cite{def2-basis-RI}.
The active orbitals include five Co $3d$ and four Co $4d$, one $\sigma$ ligand orbital (denoted as $\sigma_{3d_{xy}}$), four $\pi$ and four $\pi^\ast$ of TC-3,3, two $\pi$ and two $\pi^\ast$ of NO (see Figs. S3 to S6 in the Supporting Information for the converged active orbitals for the $S_0$, $T_1$, and $T_2$ states, respectively).
It is first seen from Table \ref{CoTN-Gap} that the present iCISCF excitation energy (20.7 kcal/mol) for the $T_1$ state differs dramatically
from that (38.6 kcal/mol) by DMRG-SCF\cite{CoTCNO3}, which indicates strongly that the two SCF calculations did not converge to the same solution.
To verify our results, we repeated the iCISCF(22,22) calculations for two triplet states (at the $T_1$ structure)
 but optimized only the second root (i.e., the weights for the first and second roots were set to zero and one, respectively).
The excitation energy of the so-obtained second root is 38.1 (or 38.8) kcal/mol if the rotations between active orbitals were included (or not).
Therefore, it can be concluded that the DMRG-SCF calculations\cite{CoTCNO3} missed the energetically lower root.
Consequently, the near-degeneracy between the $T_1$ and $T_2$ states (the lowest triplet states at two different structures\cite{CoTCNO2})
revealed by the DMRG(22,22)-SC-NEVPT2 calculations\cite{CoTCNO3} is spurious. They are actually separated by 20 kcal/mol.

\begin{figure}
\centering
\includegraphics[width=1.0\textwidth]{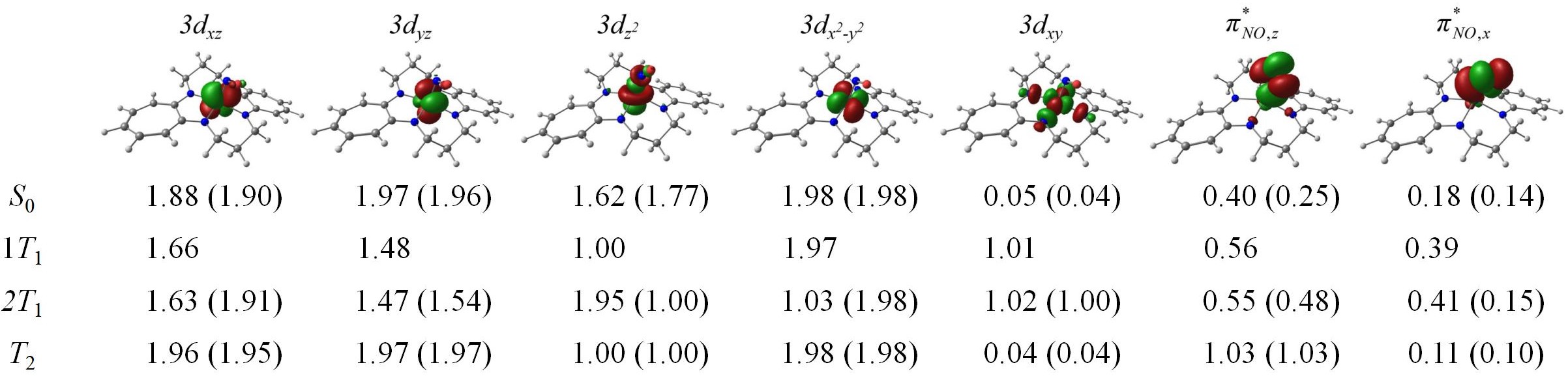}
\caption{Natural orbital occupation numbers of key active orbitals of the $S_0$, $1T_1$, $2T_1$, and $T_2$ states of [Co(TC-3,3)(NO)]. 
The $3d_{z^2}$ and $\pi^\ast_{NO,z}$ orbitals of $S_0$ are better characterized as $\pi$ bonding 
($\pi_{3d_{z^2}}$) and anti-bonding ($\pi^\ast_{3d_{z^2}}$) hybrids between $3d_{z^2}$ and $\pi^\ast_{NO,z}$
(see Fig. S3 in the Supporting Information).
Values in parentheses were obtained by DMRG-SCF\cite{CoTCNO3}, where $2T_1$ was regarded as $1T_1$.}
\label{CoTN-occ}
\end{figure}

It can be seen from  Table \ref{CoTN-CSF} that the ground state of [Co(TC-3,3)(NO)] features a closed-shell configuration $3d^6 \pi^2_{3d_{z^2}}$ 
but mixed with an anti-ferromagnetically coupled pair of half-spins at $3d^7$ ($S=1/2$) and NO radical. 
This is in line with the DMRG-SCF assignment\cite{CoTCNO3} and also supports the DFT calculations\cite{CoTCNO2}. 
The $T_2$ state is dominated by the ferromagnetic coupling between the half-spins at $3d^7$ and NO radical.
In contrast, the two triplet states ($1T_1$ and $2T_2$) at the $T_1$ structure involve more complicated spin couplings, although 
both can be characterized as a mixture of motifs $3d^7$NO and $3d^8$NO$^+$. Neither of the two is in line with the DMRG-SCF result,
as can be seen from Fig. \ref{CoTN-occ}. Therefore, the DMRG-SCF calculations should be repeated for these states.

\begin{threeparttable}
  \caption{Singlet-triplet energy gaps (in kcal/mol) of [Co(TC-3,3)(NO)] }
  \tabcolsep=20pt
  \scriptsize
  \begin{tabular}[t]{@{}ccccccccccccc@{}}
   \hline\hline
   State\tnote{a} &iCISCF\tnote{b} & SDSPT2s\tnote{b} & MS-NEVPT2s\tnote{b} & DMRG-SCF\tnote{c} & DMRG-SC-NEVPT2\tnote{c}\\
   \hline
   $1T_1$  & 20.7  & 19.9 (14.7)   & 19.7 (14.5)   &  --    &  --    \\
   $2T_1$  & 38.1  & 39.6 (34.1)   & 39.6 (34.1)   &  38.6    &  35.0    \\
   $T_2$   & 29.7  & 40.1 (35.1)   & 40.3 (35.3)   &  29.6    &  36.1    \\
   \hline \hline
  \end{tabular}\label{CoTN-Gap}
  \begin{tablenotes}
  \item[a] $1T_1$ and $2T_1$ are the lowest two triplet states at the $T_1$ structure and $T_2$ the lowest triplet state at the $T_2$ structure\cite{CoTCNO2}.
  \item[b] (22e, 22o), $C_{\text{min}}=10^{-4}$, $P_{\text{min}}=10^{-3}$, $Q_{\text{min}}=10^{-5}$, DVD; values in parentheses were derived from $\tilde{E}^{(0)}_k+E^{(2)}_k$.
  \item[c] Ref. \citenum{CoTCNO3}, where $2T_1$ was regarded as $1T_1$. 
  \end{tablenotes}
\end{threeparttable}
\vspace{1cm}

\begin{threeparttable}
  \caption{Leading configurations of the $S_0$, $1T_1$, $2T_1$, and $T_2$ states of [Co(TC-3,3)(NO)] by iCISCF(22,22)/def2-TZVP with $C_{\text{min}}=10^{-4}$ }
  \tabcolsep=25pt
  \scriptsize
  \begin{tabular}[t]{@{}llllllllcc@{}}
   \hline\hline
   State\tnote{a}& CSF\tnote{b}  &Weight (\%)  & Character of Co & Character of NO \\
   \hline
   $S_0$  & 2222000   & 55.6    & $3d^6\pi^2_{3d_{z^2}}$  ($S=0$)    &    \\
   $1T_1$ & 2uu2ud0   & 24.9    & $3d^7$ ($S=3/2$)   &  NO ($S=1/2$)  \\
          & u2u2u0d   & 11.5    & $3d^7$ ($S=3/2$)   &  NO ($S=1/2$)  \\
          & 2du2uu0   & 12.9    & $3d^7$ ($S=1/2$)   &  NO ($S=1/2$)  \\
          & d2u2u0u   &  7.3    & $3d^7$ ($S=1/2$)   &  NO ($S=1/2$)  \\
          & u2d2u0u   &  3.3    & $3d^7$ ($S=1/2$)   &  NO ($S=1/2$)  \\
          & 22u2u00   & 18.7    & $3d^8$ ($S=1$)    &   NO$^+$ ($S=0$)   \\
   $2T_1$ & 2u2uud0   & 19.4    & $3d^7$ ($S=3/2$)   &  NO ($S=1/2$)  \\
          & u22uu0d   & 12.2    & $3d^7$ ($S=3/2$)   &  NO ($S=1/2$)  \\
          & 2d2uuu0   & 10.5    & $3d^7$ ($S=1/2$)   &  NO ($S=1/2$)  \\    
          & d22uu0u   &  6.6    & $3d^7$ ($S=1/2$)   &  NO ($S=1/2$)  \\
          & 2u2duu0   &  6.4    & $3d^7$ ($S=1/2$)   &  NO ($S=1/2$)  \\
          & u22du0u   &  4.2    & $3d^7$ ($S=1/2$)   &  NO ($S=1/2$)  \\
          & 222uu00   & 16.8    & $3d^8$ ($S=1$)    &   NO$^+$ ($S=0$)   \\
   $T_2$  & 22u20u0   & 80.4    & $3d^7$ ($S=1/2$)   &  NO ($S=1/2$)  \\
   \hline \hline
  \end{tabular}\label{CoTN-CSF}
  \begin{tablenotes}
  \item[a] $1T_1$ and $2T_1$ are the lowest two triplet states at the $T_1$ structure and $T_2$ the lowest triplet state at the $T_2$ structure\cite{CoTCNO2}.
  \item[b] Sequences of active orbitals for $S_0$ and other states are Co$3d_{xz}$, Co$3d_{yz}$, $\pi_{3d_{z^2}}$, Co$3d_{x^2-y^2}$, Co$3d_{xy}$, $\pi^\ast_{3d_{z^2}}$, $\pi^\ast_{NO,x}$ [$\pi_{3d_{z^2}}$:$\mbox{ Co}3d_{z^2}\mbox{ }(24.4\%)\mbox{ and }\pi^\ast_{NO,z}\mbox{ }(74.3\%)$; $\pi^\ast_{3d_{z^2}}$:$\mbox{ Co}3d_{z^2}\mbox{ }(75.4\%)\mbox{ and }\pi^\ast_{NO,z}\mbox{ }(22.5\%)$] and Co$3d_{xz}$, Co$3d_{yz}$, Co$3d_{z^2}$, Co$3d_{x^2-y^2}$, Co$3d_{xy}$, $\pi^\ast_{NO,z}$, $\pi^\ast_{NO,x}$, respectively
      (see Figs. S3 to S6 in the Supporting Information); u: spin up; d: spin down.
  \end{tablenotes}
\end{threeparttable}

\section{Conclusions and outlook}\label{Conclusion}
It has been shown that both the computational cost and memory requirement of CASSCF-based SDSPT2 (and MS-NEVPT2) can be reduced substantially
by virtue of configuration selection,
without sacrificing the accuracy. This has been achieved by extending the highly efficient HPS-GUGA to incomplete reference spaces.
Once the reduced sub-DRTs have been constructed, all matrix elements over
internally contracted configurations can be evaluated in terms of DRT loops, precisely the same as in the case of a CAS reference.
Further combined with an integral-based prescreening of batches of those double excitations involving three active orbitals,
the resulting SDSPT2s (SDSPT2 with selection) can be applied to challenging systems that cannot be handled otherwise, as evidenced
by several examples presented here. Both SDSCI and ic-MRCISD can be simplified in the same way.
Howerver, the full potential of the reduced sub-DRT approach remains to be further explored.
For instance, the prescreening of the CI subspaces other than $\bar{\mathrm{D}}\mathrm{V}$ and $\bar{\mathrm{V}}\mathrm{D}$
has not yet been pursued. To this end, we will investigate the possibility of first constructing reduced oCFG
(instead of CSF) sub-DRTs, where the upper bounds\cite{iCIPT2} of the Hamiltonian matrix elements over oCFGs can be employed to eliminate
unimportant excited oCFGs. The resulting oCFG sub-DRTs can then be stretched to the CSF ones.
Work along this direction is being carried out at our laboratory.


\section*{Acknowledgements}
This work was supported by National Natural Science Foundation of China (Grant Nos. 22273071, 22273052 and 22373057) and
Mount Tai Climbing Program of Shandong Province.

\section*{Supporting Information}
Brief description of GUGA distinct row table; cartesian coordinates of Fe$^{\mathrm{II}}$L$_2$, [Cu$_2$O$_2$]$^{2+}$, and [Co(TC-3,3)(NO)];
active orbitals of Fe$^{\mathrm{II}}$L$_2$ and [Co(TC-3,3)(NO)];
energies of Cr$_2$ at different interatomic distances.

\section*{Conflicts of interest}
There are no conflicts to declare.

\clearpage
\newpage

\bibliography{SDSPT2}

\end{document}